\documentclass[fleqn,usenatbib]{mnras}

\usepackage{newtxtext,newtxmath}

\usepackage[T1]{fontenc}
\usepackage{subcaption}
\usepackage{multicol}
\usepackage{soul}

\DeclareRobustCommand{\VAN}[3]{#2}
\let\VANthebibliography\thebibliography
\def\thebibliography{\DeclareRobustCommand{\VAN}[3]{##3}\VANthebibliography}

\usepackage{graphicx}	
\usepackage{amsmath}	

\title{Statistical Properties of Predicted Blended Eclipsing Binary False Positives in PLATO LOPS}

\author[J. C. Bray et al.]{
J. C. Bray,$^{1,2}$\thanks{E-mail: john.bray@auckland.ac.nz}
U. Kolb$^{2}$, S. A. Mills$^{2}$
\\
$^{1}$Department of Physics, University of Auckland, Private Bag 92019, Auckland, New Zealand\\
$^{2}$The Open University, School of Physical Sciences, Walton Hall, Milton Keynes MK7 6AA\\
}

\date{Accepted XXX. Received YYY; in original form ZZZ}

\pubyear{2024}

\begin{document}
\label{firstpage}
\pagerange{\pageref{firstpage}--\pageref{lastpage}}
\maketitle

\begin{abstract}
With the planned launch of the PLAnetary Transit and Oscillation of stars (PLATO) satellite mission in 2026, an understanding of the stellar properties and spatial distribution of astrophysical false positives (FPs) is essential to ensure the limited ground-based spectroscopy resources are used efficiently to target the most likely genuine planetary transit (PT) candidates. In our previous paper, \cite{Bray_2023}, we presented the expected blended eclipsing binary false positive (BEB) percentage in the proposed Southern PLATO field, which we referred to as SPF0. This was obtained from a detailed statistical analysis of a complete synthetic rendering of SPF0. In this follow-up paper, we present a more detailed analysis of the synthetic binary population creating these BEBs including the apparent radii of the planets  mimicked, distances from Earth, orbital periods and binary component masses and evolutionary state. We examine the properties of the BEBs from 400 new independent simulations where random orbital inclinations were assigned to all synthetic binaries in the entire SPF0 region. We consider BEBs created by all eclipses of non-compact remnant binaries. Finally we examine the BEBs most likely to be mistaken for planets in the habitable zone which we define as fully or partially eclipsing binaries with periods between 180 and 1,000 days satisfying our critical inclination angle check.
\end{abstract}




\section{Introduction}
The European Space Agency (ESA) satellite mission, PLAnetary Transits and Oscillation of stars (PLATO), is currently well into the construction phase with the satellite tentatively scheduled for launch at the end of 2026. The PLATO mission \citep{rauer2024, Rauer_2014} follows other similar missions such as Kepler \citep{Borucki2010} and more recently the Transiting Exoplanet Survey Satellite (TESS) \citep{Ricker:2015aa}.
Like TESS and Kepler before it, PLATO aims to find exoplanets using the planetary transit method where a reduction in flux from the target star is detected as a planet, or planets, transit across it. Such a detection requires that the planet and the star are aligned when viewed from the satellite. Due to the low probability of this occurring, especially for small planets at large orbital distances from low-mass stars, such as Earth-like planets around Sun-type stars, a large number of target stars need to be observed if a statistically significant sample of transiting exoplanets is to be detected. As an indication, the 2014 PLATO mission proposal set a target of $>267,000$ stars to be observed for the two long duration observational phases (LOPs) \citep{Rauer_2014}. 
 PLATO is unique in that it has a very large field of view \citep[of just over 2,000 deg$^2$ exceeding even Earth2 which has a field of $\approx500$ deg$^2$][]{Earth2}, with long uninterrupted coverage, and its priority targets are very bright stars ($m_v<13$). Ground-based radial velocity and spectroscopic analysis providing essential information for confirming candidates and characterising planets will be possible on a larger number of host stars with longer-period planets than has been the case for Kepler or TESS.
  
The current proposal for the LOPs is to observe two fields of about 2,132 deg$^2$, one located in the Northern Hemisphere (LOPN), and the other in the Southern Hemisphere (LOPS), each continuously for a duration of two years. However, the final observing strategy has yet to be confirmed.  The two-year minimum observation period is designed to fulfill one of PLATO's primary goals, which is to find Earth-like planets \citep{rauer2024, Rauer2018}. 
   
With such a large field of view (FOV), and short focal length, a large pixel scale is required to cover the proposed fields. For PLATO this translates to an area of 15 by 15 arc seconds (15\arcsec $\times$ 15\arcsec) per pixel. In addition, to reach the large number of target stars required, the observational fields must be located in regions with a relatively high stellar density. Logically, the denser the observational field the higher the likelihood of astrophysical false positives from blended eclipsing binaries, since the foreground and background fields are also proportionately more dense. 
   
In this work, we predominantly use the original proposed southern PLATO field, which was referred to as SPF0, (with field centre Galactic co-ordinates of $l = 253^\circ$ and $b = -30^\circ$), as a case study to quantify the dependency of false positives (FPs) from eclipsing binaries blended in the same analysis area as the target stars, on variables such as planetary radius (as measured from the transit depth), distance to the eclipsing binary causing the FP, eclipse type (full eclipse, deep grazing eclipse, shallow grazing eclipse) and orbital period. While we recognise that other phenomena, such as stellar activity, can also result in FP planetary transit signals, in this research we focus solely on FPs resulting from blended eclipsing binaries. This type of FP we refer to hereafter as a BEB (blended eclipsing binary FP).

The location of the first long pointing field located in the Southern Hemisphere was finalised in mid-2023 (LOPS2), and is centred on $l = 255.9375^{\circ}$ and $b = -24.62432^{\circ}$ \citep{nascimbeni2025}. 
With a centre longitude of $l=255.9375^{\circ}$ versus the original proposed field of $l=253^{\circ}$, this is located closer to the Galactic centre. More importantly, with a latitude of $b = -24.62432^{\circ}$ versus the original proposed field of $b=-30^{\circ}$, the field centre is $\approx5.4^{\circ}$ closer to the Galactic plane, dropping the lower portion of the field into the plane itself. 

In our first paper \citep[][hereafter Paper 1]{Bray_2023}, we referred to both the original proposed long pointing field and our synthetic rendering of that field as the Southern PLATO field SPF0. The naming convention for the long pointing fields was subsequently changed (from SPF to LOP), and so to eliminate any confusion and to ensure this paper uses the same naming   convention as the PLATO consortium, we refer to our synthetic rendering of the original SPF0 field as SYNTH0 and to the original proposed PLATO (SPF) field as LOPS1.
 

This paper is set out as follows: in Section \ref{tpm} we present a brief description of the PLATO mission relevant to our research. In Section \ref{pst} we briefly describe the stellar evolution, stellar and planet population synthesis, planet transit (PT), and FP analysis tools used to assess the expected BEB occurrences. In Section \ref{targets} we compare our PLATO P5 target star selection with the actual P5 target stars identified in the PLATO input catalogue (PIC), PIC1.1.0 and the latest release of the PIC, PIC2.0.0. In Section \ref{planets} we compare our projected planet yield with more recent and more sophisticated estimates to ensure that our planet yields (and hence \%BEB results), are plausible. In Section \ref{seba} we define our eclipse types and our BEB analysis method in more detail. In Section \ref{gpebc1} we compare our results with those in \cite{Bray_2023} and discuss the transit duration check and its impact on our results 
and in Section \ref{fpdisc} we examine the properties of BEBs mimicking planets with `Earth-like' periods between 180 and 1,000 days, and finally in Section \ref{dc} we discuss our findings and present our conclusions.

\section{The PLATO Observing Strategy and Target Group Selection}\label{tpm}
Below we outline the details of the PLATO mission relevant to our research, for a more comprehensive account of the mission and its aims refer to \cite{rauer2024, Rauer_2014, Rauer2018} and references therein. 

The PLATO mission is a four-year mission comprising two proposed LOPs, one in the Southern Hemisphere and one in the Northern Hemisphere. At this point it is intended that each of the LOPs will have a two-year duration. If the mission is extended, it is envisaged that a number of `step-and-stare' phases will be added, covering additional areas of sky but for much shorter time intervals, most likely 2-5 months each. The proposed LOP fields are $\approx2,232$ deg$^2$ with both constrained to have their centres located within spherical caps with the ecliptic coordinate $|\beta| > 63^{\circ}$ \citep{Nascibeni2022}. While the location of the Northern Hemisphere LOP is still being assessed, the location of the Southern Hemisphere LOP field has been confirmed to be centred on $l \approx 255.9^{\circ}$ and $b \approx -24.6^{\circ}$ \citep{nascimbeni2025}. The current working assumption within the consortium is that the Northern Hemisphere LOP field will be centred on $l \approx 81.6^{\circ}$ and $b \approx 24.6^{\circ}$.

The largest PLATO target star group P5, is defined as stars with magnitude m$_v\leqslant13.0$ of type F5 to late K with a maximum effective temperature ($T_{\textrm{eff}}$), of 6775K and a minimum of 3875K \citep{PLATORedBook}. For this research we concentrate on the P5 target star group contained in LOPS1. This location and target star group have been selected as they provide the largest number of stars and hence will produce the most statistically significant BEB results. 

\section{Stellar evolution, planet population synthesis and resulting BEB and planetary transit (PT) detections}\label{pst}
In this work, as in Paper 1, we utilise the Binary Stellar Evolution and Population Synthesis (\textsc{BiSEPS}), stellar models to create synthetic single and binary star populations for LOPS1 in its entirety. We refer to this rendering as SYNTH0. 
We estimate the PTs and BEBs as outlined below and then combine these to provide a statistical estimate of the \%BEB which we define as (BEB/(BEB+PT)) $\times\ 100$.

We highlight that our analysis is based entirely on our synthetic rendering of the LOPS1 field and we do not use data from any survey catalogs.
  
\subsection{Stellar Evolution - The \textsc{BiSEPS} code}
\textsc{BiSEPS} is a population synthesis tool developed by \cite{Willems_2002}. It uses large libraries of different metallicity single and binary stellar models to create Galactic populations. The stellar models are created using simplified binary evolution algorithms based on the prescriptions of \cite{Hurley_2000,Hurley_2002}. The algorithms consider mass loss from stellar winds, angular momentum loss via gravitational waves, magnetic braking and Roche-lobe overflow. The code has been utilised in a number of stellar population studies \citep[e.g.][]{Willems_2004, Willems_2006, PL-Farmer_2013}. Readers wanting more detail on the code should refer to the above studies.

The initial mass parameter space for the primary stars ($M_1$), is divided into 50 evenly spaced logarithmic bins between 0.1M$_{\odot}$ and 20M$_{\odot}$ and into 300 evenly spaced logarithmic orbital separation bins between 3R$_{\odot}$ and $10^6$R$_{\odot}$. The secondary stars ($M_2$), are chosen from the same parameter space as $M_1$ but only systems where $M_1 > M_2$ are evolved. Initial orbits are assumed to be circular and subsequent orbits are kept circularised at each time step. 

For practical reasons, single stars are modelled as binary systems with an orbital separation of $10^7$R$_{\odot}$. The primary single star mass parameter space ($M_1$) is divided into 10,000 evenly spaced logarithmic bins between 0.1M$_{\odot}$ and 20M$_{\odot}$ with the secondary star mass ($M_2$), set at 0.1M$_{\odot}$.

\subsection{Population Synthesis}
Population synthesis in \textsc{BiSEPS} is carried out by randomly selecting single and binary models from the libraries created above. We assume a binary fraction of 0.5 and assign probabilities for each model using $M_1$ and an initial mass function (IMF) following \cite{KIMF} with $dN/dM \propto M^\Gamma$ as follows;\\

$\Gamma = 
\left \{
    \begin{array}{l l}
    -1.3  \text{ for} & M_1<0.5\text{M}_{\odot}\\
    \\
    -2.2  \text{ for} & 0.5\leqslant M_1<1.0\text{M}_{\odot}\\ 
    \\
    -2.7 \text{ for} & 1.0\text{M}_{\odot}\leqslant M_1\\        
    \end{array}
\right.$\\

We create a synthetic rendering of the Milky Way  assuming a thin disc with a metallicity of ${Z}=0.02$ embedded in a thick disc with a metallicity of ${Z}=0.0033$. The stellar density of both discs is modelled assuming the double exponential distribution outlined below;
\begin{equation}
\Omega(R_g,z)=\frac{1}{4\pi h^{2}_{r}h_{z}} exp\left(\frac{-R_g}{h_{r}}\right)exp\left(\frac{-|z|}{h_{z}}\right)
\end{equation}
with $h_{r}$ = 2.8kpc and $h_{z}$ = 300pc for the inner disc and $h_{r}$ = 3.7kpc and $h_{z}$ = 1kpc for the outer disc. The Sun is located at $R_\textrm{g} = 8.5$kpc and $z = $30pc. 
Star formation is assumed to occur in the thick disc for the first 3Gyr and in the thin disc from 3Gyr to 15Gyr. 

\subsection{Eclipsing binary calibration and BEB detection}
We used the \textsc{BiSEPS} models created by \cite{PL-Farmer_2013} which were modified to reproduce the Kepler input catalogue (KIC) from \cite{Brown_2011} in $\log(g)$ vs $\log(T_{\textrm{eff}}$) space, when subjected to the Kepler Stellar Classification program. For more details on the code used to create these models readers are directed to \cite{PL-Farmer_2013}.

The synthetic mass ratio distribution of eclipsing binaries detectable by Kepler increases with decreasing period and matches the distribution of the Kepler Eclipsing Binary Catalog \cite[KEBC : ][]{Kirk_2016}, satisfactorily for periods larger than about 40 days. Below about 10 days the KEBC mass ratio and period distributions flatten off while the synthetic distributions continue to rise, resulting in a model over-prediction by about a factor of two for these short-period binaries. In contrast, for all mass ratio binaries with periods between 10 and 40 days, the synthetic distribution slightly under-predicts the population.


We address this by introducing two weighting factors in the synthetic sample, one for the mass ratio and a second for the period.  
For systems with $P<10$days, we remove the mass ratio discrepancy by introducing a calibration weighting which is proportional to $q^{2.5} \times f(\log P)$ where $f(\log P)$ increases effectively linearly from 0.4 at $P=1$day to 1 at $P=10$days. Then for all systems with $P<40$days a second, minor weighting factor is applied (calculated iteratively), to match the synthetic short period binary distribution to the Kepler observations \cite[see][for more details]{PRThesis}.

For the BEB analysis, we identify all single stars matching the P5 target criteria in SYNTH0 and carry out 400 simulations on the binary populations surrounding each target. To approximate the multi-pixel mask suggested by \cite{Marchiori:2019aa} we analyse an area of 30\arcsec x 30\arcsec (four PLATO pixels), centred on each P5 target star (see Section \ref{seba} for details). For the identified binary populations, we assign random inclination angles and then identify all binary eclipses resulting in a flux drop greater than our calculated PLATO instrument noise (see Paper 1 for the noise calculation). For our flux-drop calculations, we use an out-of-transit flux equal to the total flux from all stellar objects in the 30\arcsec x 30\arcsec area. We then calculate the apparent radius of the planet mimic from the flux drop and record the period.


\subsection{Synthesis and detection of the PLATO P5 exoplanet population}\label{ied}
Data from microlensing surveys suggests there is at least one planet per star \citep{Cassan_2012}, so as a first-order estimate, we identify all synthetic single stars in SYNTH0 meeting the P5 target star criteria and seed each with a single planet and assign a planet radius ($R$) and orbital period ($P$). 

To obtain realistic $P$ and $R$ distributions for these planets, we calibrate our $P$ and $R$ distributions using the Kepler field data. We do this by initially seeding each single star in a synthetic rendering of the Kepler field with one exoplanet, randomly chosen from an initially flat distribution in $P$ and $R$. Each system is then assigned a random orbital inclination angle. Mimicking the Kepler detector characteristics, we extracted the detectable synthetic exoplanet distribution over $P$ and $R$ and compared this against the ‘observed’ exoplanet distribution taken from the NASA Exoplanet Science Institute (NExScI) catalogue of confirmed Kepler planets. The seeding probabilities were then iteratively adjusted until a satisfactory match between the synthetic and real exoplanet distributions was achieved. While it could be argued that the comparison of our detected planets with the Kepler planet `candidates' may have been more appropriate, in 2018 when the project commenced, it was felt that `confirmed' planets would be a more reliable dataset minimising contamination from potential FPs. In addition, a comparison of the distributions for $P$ and $R$ between the confirmed planets and the confirmed and candidate planets combined showed that the $P$ and $R$ distributions were broadly similar for  $\log(R/$R$_{\oplus}) > 0$ which includes the planetary radius range of most interest to PLATO. We note that the Kepler planet catalogue is least complete at long periods and small planet radii so model results for $\log(R/$R$_{\oplus}) < 0$ are likely to have higher uncertainties and should be interpreted in this context  
\citep[see Section~3.4 in][for further detail] {Bray_2023}. 

We then seed each synthetic P5 target star with a single planet with a period and radius selected from the distributions calculated above. We carry out 200 simulations where we assign random planetary inclination angles to each planet and identify all planets with transits resulting in a flux drop which is greater than our calculated instrument noise (see Paper 1 for the noise calculation). As for the BEBs, the out-of-transit flux for the PTs is calculated from the total flux from all stellar objects in the 30\arcsec x 30\arcsec areas. We then calculate the planet radius from the transit depth and record the orbital period.

There has been considerable research on exoplanet system properties in recent years and we recognise that our one planet per star and planet period/radius selection method is somewhat simplistic. In Section~\ref{planets} we demonstrate that our model nonetheless performs satisfactorily for our purpose. A more sophisticated planet seeding/selection method is beyond the scope of this research but as we provide planet and false positive numbers separately the reader may use their own planet statistics to determine a relative false positive rate on the basis of their preferred planet model and our BEB numbers. 

 
\section{Synthetic vs PIC1.1.0 and PIC2.0.0 Target Star Population}\label{targets}
Before any detailed analysis of the BEBs is attempted, we wanted to verify that our SYNTH0 target star population provides a good representation of the target stars identified in the PIC, particularly in their magnitude distribution. Since our BEB analysis is statistical, any systematic error in the target star magnitude distribution will result in a bias in the BEB distribution. For instance, if our target star magnitude distribution is biased toward brighter stars, our BEB distribution will be biased towards smaller planets as the same eclipses after dilution with the flux from the target star analysis area will result in smaller flux reductions. While the BEB statistics are not directly affected by the target star distance distribution (assuming the target star magnitude distribution is correct), the binary systems that eclipse, and result in BEBs, are selected from the same galactic density model as our synthetic target stars and any systematic error in the target star distance distribution will result in a corresponding distance distribution error in the background binary and hence BEB population, which will in turn introduce a bias in the BEB statistics. For example, if our binary star distances are systematically underestimated, our BEBs statistics will overestimate the number of larger planet mimics in our BEB population. 


The PLATO consortium created the original PLATO input catalogue PIC1.0.0 \citep[][]{Montalto2021} using data predominantly from Gaia \citep{2016Gaia} DR2 (RD05). We used PICtarget110 from PIC1.1.0 which identifies target stars in both the proposed Northern, and Southern, PLATO LOP fields, however, in September 2023 a major update of the PIC was released (PIC2.0.0). This new PIC is based on Gaia DR3 \citep{2023GaiaDR3} and contains pLOPS2PICtarget2.0.0.1. (hereafter referred to as PICtarget200), which supersedes the LOPS1 field in PICtarget110 \citep{Nascibeni2022}. 


As well as being based on more recent Gaia data, PICtarget200 covers the confirmed LOPS2 field which is located closer to the Galactic centre and plane. As a result, PICtarget200 contains significantly more objects. However, we expect the target star distributions, such as distance and magnitude to be relatively unchanged. This is evident in the similarity of the distributions shown in Figures \ref{fig:PICsynthDists} and \ref{fig:Gmaglog200}. This similarity in distributions is expected because the target stars are bright and have a relatively narrow temperature range, so while looking at a more dense field will slightly increase the number of stars meeting this criteria, the temperature requirement  does not allow for hotter, more distant stars to be included.





We reiterate that our synthetic field (SYNTH0), has been created to provide a statistical representation of the characteristics of the stellar objects, and hence BEBs, in LOPS1. As a result it is not expected to accurately reproduce the specifics of individual objects in PICtarget110 or PICtarget200. However, as highlighted above, we expect a relatively good correlation between the SYNTH0 P5 target population and both PIC110 and PIC200 in terms of numbers, distance and magnitude distribution. 

To test this, we carry out a number of comparisons between our target star population in SYNTH0 and the two versions of the PIC.

\subsection{The number of target stars} 
Our SYNTH0 target dataset contains 78,257 single stars matching the P5 criteria outlined in Section \ref{tpm}. This compares with 140,105 objects contained in PICtarget110, and 167,103 in PICtarget200 using the same selection criteria. 

We believe this discrepancy is mostly likely due to the inability of Gaia (on which the PIC is based), to resolve very close binary systems. The resolution of Gaia is stated as 0.23\arcsec\  prior to on-ground processing \citep{Bruijne2015} and 0.1\arcsec\ after on-ground processing \citep{Harrison2011}. To test this assumption we identified SYNTH0 background binary systems matching the P5 criteria that have a separation of 0.1\arcsec\ or less. We find 49,779 binary systems which fit this definition, bringing the apparent synthetic `target' star total to 128,036. This reduces the target number discrepancy to less than 10 percent which we believe represents an acceptable agreement with PICtarget110 target star numbers. 

We note that there is no mention in \cite{Nascibeni2022} or \cite{nascimbeni2025} of the use of the Gaia Renormalized Unit Weight Error (RUWE) to reduce the likelihood of binary stars in the Gaia catalogue being identified as single stars and we did not utilise the Gaia RUWE in our interrogation of the Gaia catalogue. 
 
\subsection{Distance distribution of target stars} 
\begin{figure}
\centering
   \includegraphics[width=\hsize]{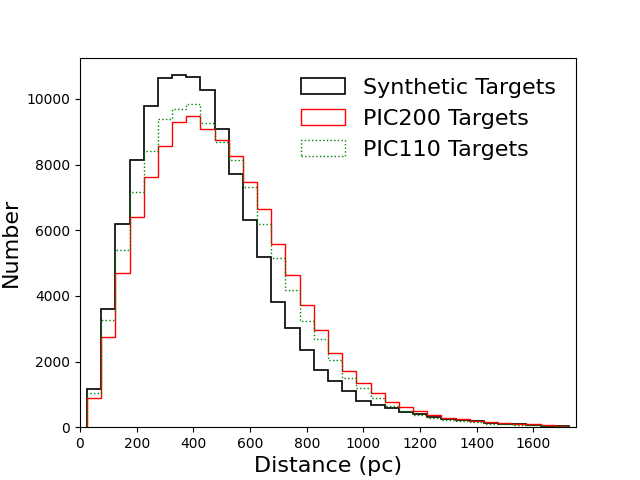}
      \caption{Distribution of distances to PLATO P5 target stars in the synthetic southern PLATO field SYNTH0, versus the distance distribution in PICtarget110 and PICtarget200. The number of objects in PICtarget110 and PICtarget200 is normalised to the numbers in SYNTH0.}
         \label{fig:PICsynthDists}
\end{figure}

In Figure \ref{fig:PICsynthDists}, we show the PLATO P5 target star distance distribution of PICtarget110, PICtarget200 and  SYNTH0. The peak probability for each distribution occurs at $\sim$380pc although the height of the peak is slightly lower for PICtarget110 and PICtarget200 due to their higher weighting for objects with distances between 580pc and 1,100pc. PICtarget110 also contains $\sim 100$ objects more distant than 2,100pc out of a total of 140,105 (less than 0.1 percent), which our SYNTH0 distribution does not. Such a low number of additional distance targets is not expected to have any significant effect on our findings. Overall we deem the distance distribution of our synthetic population and PICtarget110 and PICtarget200 to be in sufficiently good agreement for the purpose of our analysis. 

\subsection{Magnitude distribution of target stars}
\begin{figure}
\centering
   \includegraphics[width=\hsize]{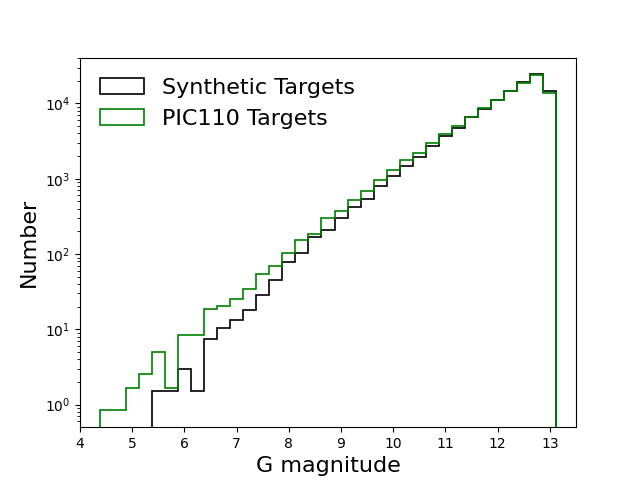}
      \caption{Distribution of Gaia magnitudes ($G_\textrm{mag}$), for PLATO P5 target stars in the synthetic southern PLATO field SYNTH0, versus the magnitude distribution in PICtarget110. The number of objects in PICtarget110 is normalised to the numbers in SYNTH0. The y-axis is a log scale.}
         \label{fig:Gmaglog110}
\end{figure}

\begin{figure}
\centering
   \includegraphics[width=\hsize]{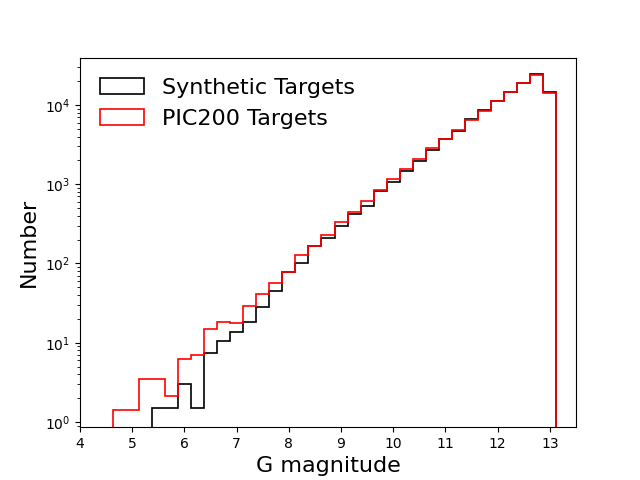}
      \caption{Distribution of Gaia magnitudes ($G_\textrm{mag}$), for PLATO P5 target stars in the synthetic southern PLATO field SYNTH0, versus the magnitude distribution in  PICtarget200. The number of objects in PICtarget200 is normalised to the numbers in SYNTH0. The y-axis is a log scale.}
         \label{fig:Gmaglog200}
\end{figure}

Next we consider the distribution of the target stars' Gaia magnitude ($G_\textrm{mag}$). As shown in Figure \ref{fig:Gmaglog110}  we find a good agreement between PICtarget110 and our SYNTH0 target dataset. We see an even better correlation between PICtarget200 and SYNTH0 (see Figure \ref{fig:Gmaglog200}).

While the number of bright stars is the same in both PICtarget110 and PICtarget200, PICtarget200 has more very faint stars and as a result, when normalised, PICtarget200 more closely resembles our synthetic target dataset as evident in the 7-9 magnitude range.


 

\section{Projected planet yield from the synthetic population}\label{planets}
As in \cite{Bray_2023}, we again create an intrinsic exoplanet population by seeding every single star with a single planet as outlined in Section \ref{ied}.

Much more sophisticated methods utilising planet formation models to determine the planetary period and radius distributions, have been carried out by other PLATO consortium authors e.g \cite{Matuszewski2023M} and \cite{Heller2022}. As a cross-check on our simplistic planet period and radius distributions we compare our projected radii and orbital periods to those in \cite{Matuszewski2023M}. Our synthetic field generally predicts fewer planets than the \cite{Matuszewski2023M} analysis for 2+2 year observing runs, and as a result our detection numbers are most similar to those of \cite{Kunimoto2020} (hereafter referred to as KM20). 

\begin{figure}
\centering
   \includegraphics[width=\hsize]{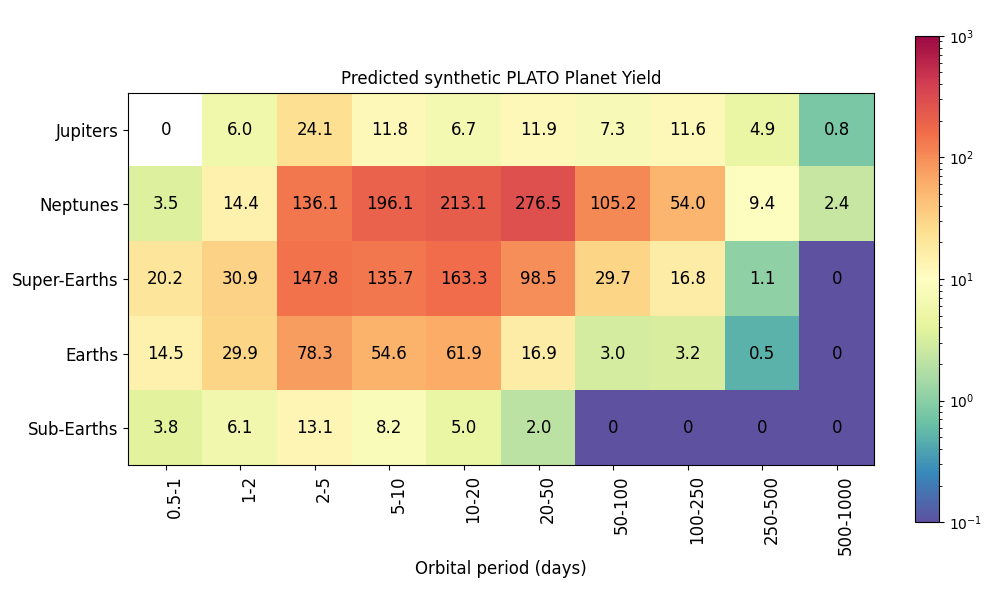}
      \caption{Projected planet yield by orbital period and planet radius from SYNTH0 binned to match that of \protect\cite{Kunimoto2020}.}
         \label{fig:plyield}
\end{figure}

In Figure \ref{fig:plyield} we show our planet detection numbers in SYNTH0 for the same planet radius and orbital period bins as KM20 in \cite{Matuszewski2023M} Figure 7.

\begin{figure}
\centering
   \includegraphics[width=\hsize]{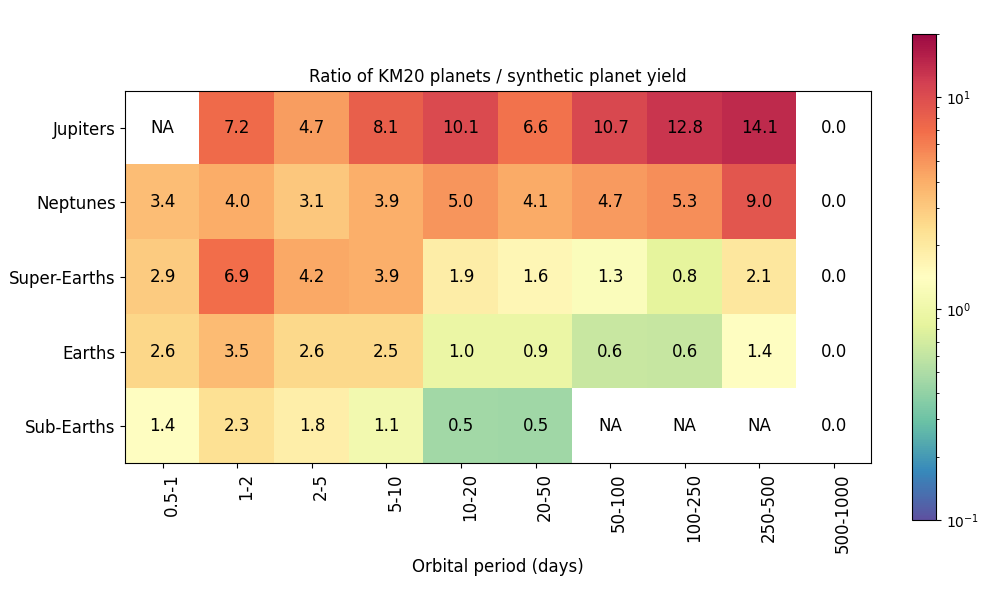}
      \caption{Ratio of \protect\cite{Kunimoto2020} projected planet yield to SYNTH0 planet yield by orbital period and planet radius}
         \label{fig:plratio}
\end{figure} 

We find our planet yield is in good agreement with KM20 in terms of overall distribution shape, with the peak number of planets occurring in the Neptune size planets with periods between two and 100 days and the second highest numbers for Super-Earths, with periods between two and 50 days. As for the total number, for planets up to Neptune radii, for all periods, we find our synthetic planet detections total $\sim40$ percent of those cited in KM20. As our analysis was focused on Earth-like planets we do not discuss Neptune and Jupiter type planets other than to note that our planet number under-estimation seems to increase as the planet radius increases.

In Figure \ref{fig:plratio} we show the ratio of the KM20 projected planet yield to the \textsc{BiSEPS} LOPS1 synthetic planet yield in SYNTH0, by orbital period and planetary radius. The ratio is shown as `NA' if the synthetic planet yield was less than $1 \times 10^{-3}$ and `0' if the KM20 planet yield in that cell was zero. For planet radii in the Super-Earth range and smaller, we find a maximum ratio of $\sim 7$. For planets with periods between 10 and 1,000 days in this planet range, incorporating Earth-like planets, we find that the two models are in agreement within a factor of $\sim 2$. We reiterate that while we believe our most robust statistic is the \%BEB, the correlation in the planet range of Earth-like planets is reassuring.


With our target star numbers, distances and magnitudes in SYNTH0 showing satisfactory consistency with both PICtarget110 and PICtarget200 and our planet number, period and radii distributions broadly consistent with \cite{Kunimoto2020}, we now consider the properties of the false positives generated by BEBs in SYNTH0.

\section{Eclipse definitions and BEB Analysis}\label{seba}


The work of \cite{Marchiori:2019aa} suggests that the best balance between maximising planet yield and minimising the  false positives is achieved with a binary mask analysing the flux over a number of pixels around the target star rather than just the single PLATO pixel containing the target star. While in reality the number and arrangement of PLATO pixels included in the mask varies, we have approximated this mask by analysing each target star over an area of 30\arcsec x 30\arcsec" i.e. areas of two by two PLATO pixels which we refer to as `super-pixels'. 

For the PT analysis, we select every single star in the SYNTH0 field meeting the PLATO P5 target star criteria outlined in Section \ref{tpm} and seed each with a single planet with a period and radius selected from the distributions calculated in Section \ref{ied}. We then assign each system a random orbital inclination angle, and
if the orbital inclination angle exceeds the critical orbital inclination angle, signifying a transit, the planet radius is calculated using the flux drop during the transit. The out-of-transit flux is taken as the total flux in the four-pixel area and since $R_{\textrm{Planet}} << R_{\textrm{Target}}$ the critical angle for a PT is calculated assuming the planet as a point. A PT event is recorded if the blended flux reduction exceeds the estimated detector noise. (See Paper 1 for details on how this estimated detector noise is calculated).




For the BEB analysis, we select, from the \textsc{BiSEPS} master population files, all binary systems within a square of 30\arcsec x 30\arcsec (four PLATO pixels), centred on each PLATO P5 target star.


As with the target stars, we assign a random orbital inclination angle for each binary system identified and, if the orbital inclination angle exceeds the critical orbital inclination angle, signifying an eclipse, the primary and secondary eclipse depths are calculated taking the total flux from the super-pixel as the out-of-transit flux. For BEBs, only the deepest eclipse is selected and the apparent radius of the planet mimic is again calculated using the flux drop during eclipse. If the resulting blended flux reduction exceeds the estimated detector noise, a BEB event is recorded. 

\cite{Sullivan2015} point out that the primary and secondary eclipse depths of binaries with mass ratios close to one could be indistinguishable, leading to a false positive planet detection with a period of half of the eclipsing binary period. We find that if all BEBs with mass ratios greater than 0.8 were mis-identified as planet transits with half the binary period we would undercount BEBs in the period range of interest, $180<P_\textrm{orb}(\textrm{days})<1,000$, by 1,616. Conversely, adding in all BEB eclipsing binaries with mass ratios greater than 0.8 with $1,000<P_\textrm{orb}(\textrm{days})<2,000$ increases the number of BEBs in the period range of interest by 722. While these changes may affect the relative planet radius distribution, the effect on the BEB numbers is a reduction of approximately 11 percent (i.e. a net reduction of 894 from 8,344). Hence, from this perspective, our BEB numbers can be regarded as a `worst case' scenario.


The eclipse categories we use in our analysis are defined as follows; in each case the eclipsing binary is located within a 30\arcsec\ box centred on a target star: 
\begin{itemize}
    \item [$\bullet$] Compact remnant - all eclipses: The eclipsing  object is a compact remnant. We include both full and grazing eclipses by compact remnants. We record 4,402 eclipses of this type in 400 simulations.
    \item [$\bullet$] Stars - full eclipses: The eclipsing star is any stellar type, other than a compact remnant. Only full eclipses are recorded, i.e. the entirety of the eclipsing star must enter the disc of its companion at some point in the eclipse. We record a total of 190,232 eclipses of this type in 400 simulations.
    \item [$\bullet$] Stars - deep grazing eclipses: The eclipsing star is any stellar type, other than a compact remnant. At some point during the eclipse the centre of the eclipsing star enters the disc of its companion, but not the entirety of the eclipsing star's disc. We record a total of 145,468 eclipses of this type in 400 simulations.
    \item [$\bullet$] Stars - shallow grazing eclipses: The eclipsing star is any stellar type, other than a compact remnant. During the eclipse part of the companion is obscured by the star but the centre of the eclipsing star does not enter the disc of its companion at any point. We record a total of 77,249 eclipses in 400 simulations.
\end{itemize}




In Paper 1 we excluded grazing eclipses from our analysis, assuming that in the PLATO data a `V' shaped light curve would be identified and the grazing binary eclipse would be discounted as a planet transit. However at low signal to noise ratios grazing eclipses  may not be readily identifiable by this method.


A more detailed appraisal as to which grazing eclipses may not be discerned as `V' shaped is beyond the scope of this work. 
Instead, we adopt a conservative or `worst case' scenario, adding both deep and shallow grazing eclipses to the full eclipse numbers. 



PTs and BEBs with apparent radii between $-0.4 \leqslant \log(R/$R$_{\oplus}) \leqslant 1.8$ were binned in radius bins of width $\log(R/$R$_{\oplus})=0.2$. We restrict our binary periods to the range $-0.4\leqslant \log(P)<3.2$ (days), which we also sort into bins of width $\log(P)=0.2$. While PLATO's focus is on Earth-like planets around Sun-type stars we still wish to identify trends in the BEBs by planet radius, hence here we show results for planets with radii of $\log(R/$R$_{\oplus})$ up to $1.2$.


Since all of our binary models have circular orbits, and we consider only the deepest eclipse, we ignore the possibility of secondary only eclipses as a result of elliptical orbits. The probability of the precise alignment required for these events to be observed is less than 0.1 percent \citep{Santerne2013} and so their omission will have no material effect on the results presented here.


\section{Comparison to Paper 1}\label{gpebc1}
Each simulation is independent and randomly allocates the binary orbital inclination angle according to an isotropic distribution across solid angle. 
As a result we expect to see some statistical variations when we compare different simulation sets. Comparing the average number of BEBs for the P5 target group from this work to those found in Paper 1, we find a good match with $\sim159$ BEBs per simulation in 15 simulations carried out in Paper 1 vs $\sim170$ per simulation for the 400 simulations carried out in this work. 


For each identified BEB we use Equation \ref{Eqn2} to calculate the apparent transit duration,
\begin{equation}
T_\textrm{dur}=\left(\frac{P_\textrm{Bin}} {\pi}\right) \times \arcsin\left(\frac{\sqrt{({R_1 + R_2)^2 - (a_\textrm{Bin}^2 \cos^2{i_\textrm{Bin}}})}}{a_\textrm{Bin}}\right),
\label{Eqn2}
\end{equation}
where $P_\textrm{Bin}$ is the binary orbital period, $a_\textrm{Bin}$ is the calculated orbital separation, $R_1$ and $R_2$ are the binary star radii and $i_\textrm{Bin}$ is the inclination angle of the binary system.

Using the period of the BEB and Kepler's third law, we calculate $a_{\star}$, the separation of the apparent planetary system, assuming the planetary system mass is $\approx M_{\star}$ (i.e. $M_{\star} >> M_\textrm{Planet}$). We then use Equation \ref{Eqn22} to calculate $i_{\star}$, the inclination angle of the apparent planetary system required to produce this transit duration,



\begin{equation}
T_\textrm{dur}=\left(\frac{P_\textrm{Bin}} {\pi}\right) \times \arcsin\left(\frac{\sqrt{({R_{\star} + R_\textrm{Planet})^2 - (a_{\star}^2 \cos^2{i_\star}})}}{a_{\star}}\right).
\label{Eqn22}
\end{equation}

Using the radius of the target star and the apparent planet, we then calculate the critical inclination angle for a transit of the mimicked system to occur. We then remove any systems where the inclination angle calculated in Equation \ref{Eqn22} is smaller than the critical inclination angle for this target/planet combination.

\begin{figure}
\centering
   \includegraphics[width=\hsize]{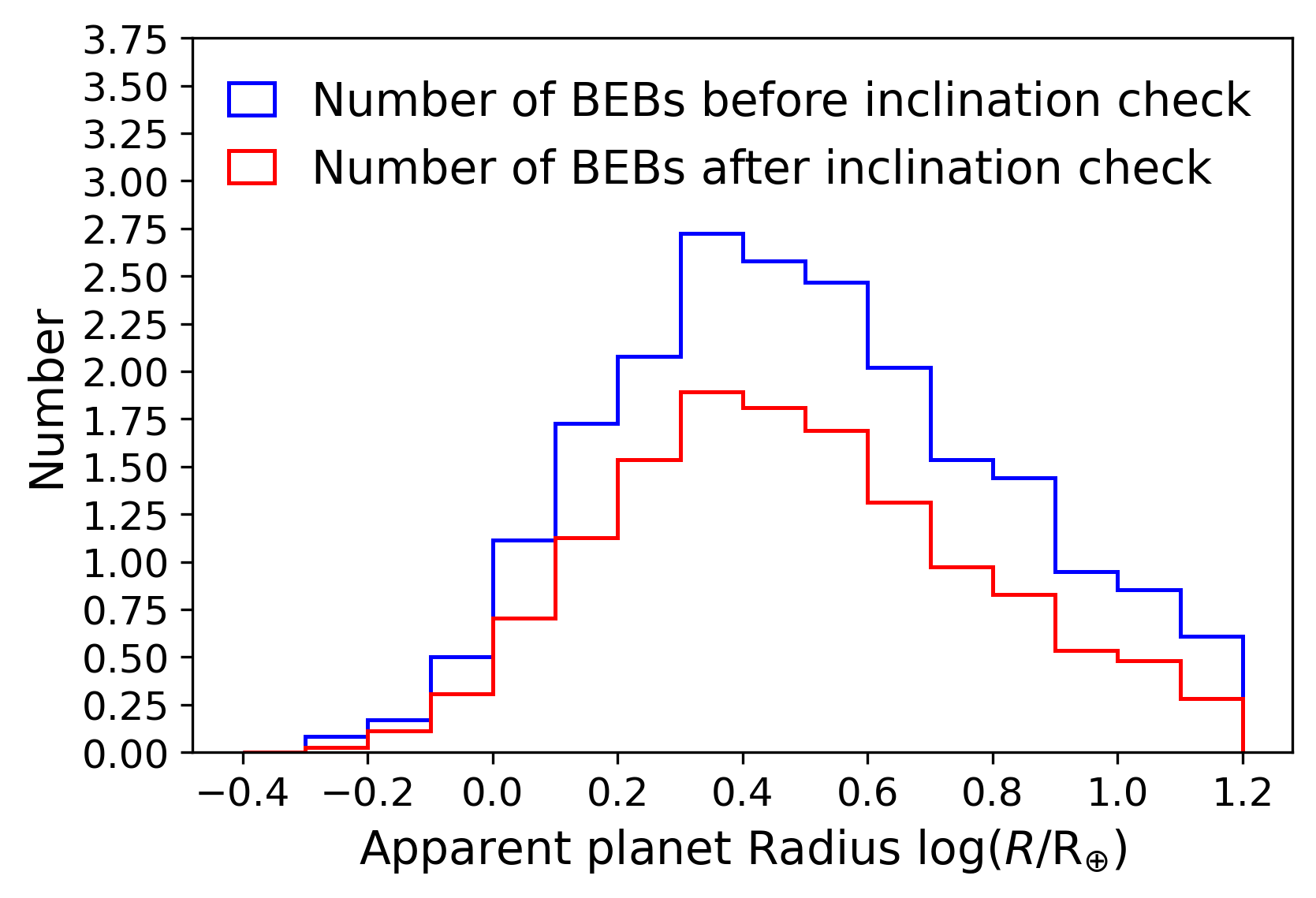}
      \caption{Number of false positives from blended eclipsing binaries (BEBs) in SYNTH0 before and after the orbital inclination angle check. 
      }
\label{fig:FPPC}
\end{figure}

We find that this additional vetting criterion reduces the BEB numbers by $\approx 30$ percent (see Figure~\ref{fig:FPPC}). 
If we exclude BEBs that fail the inclination angle test we find the number of BEBs mimicking planets with radii $-0.4 \leqslant \log(R/$R$_{\oplus}) \leqslant 1.2$ and periods $180<P_\textrm{orb}<1,000$ (days) reduces from $\sim21$ to $\sim14$.

\section{False positive blended eclipsing binaries with Earth-like periods}\label{fpdisc}
 Since the P5 targets have a stellar temperature range of $3,875$K$< T_\textrm{eff}<6,775$K, the Earth-like (EL), planet period/stellar temperature group will include planets with shorter periods than Earth orbiting lower temperature stars than the Sun as well as longer period planets orbiting stars hotter than the Sun. As a result, we choose a wide EL period range of 180 to 1,000 days. While this range is somewhat arbitrary, it is deliberately broad to ensure no potential candidates of interest are excluded. From Figure \ref{fig:m1} onwards, we show only those eclipses whose period is `Earth-like' i.e. between 180 and 1,000 days and that pass the orbital inclination check. %
 In the following figures 
 the histogram is calculated for each of the 400 simulations and the mean across those simulations is shown. The error bars express the mean error for each point, i.e.\ the standard deviation of the 400 values of that point divided by the square root of the number of simulations. A total of 400 simulations have been used as this reduces the mean error sufficiently to show the features described in the following figures.

 We highlight that using only two metallicities to model the Milky Way, one for the thin disc of $Z=0.02$ and one for the thick disc of $Z=0.0033$, rather than a progressive metallicity distribution, has resulted in a somewhat exaggerated reduction in the primary mass distribution at masses greater than $\approx 0.9$M$_{\odot}$ (evident in the upper panels of Figures \ref{fig:m1} to \ref{fig:m3}). The binary stars in the thick disc (sub-solar metallicity) are more evolved than those in the thin disc hence, there are no high mass stars and so no binaries containing stars with mass larger than about 1M$_\odot$ to create BEBs and mimic planets. While the Galactic metallicity evolves with age, the fact that we model the Galaxy using a step function for the metallicity evolution means the effect of the population transition is exaggerated. As a result, while the trends we identify are present, some caution is advised in interpreting the sharp transitions shown in Figures \ref{fig:m1} to \ref{fig:m3}.
 
 
\subsection{Masses of BEBs by eclipse type, planet mimic radius and metallicity}
\begin{figure}
\centering
   \includegraphics[width=0.7\textwidth, height=0.55\textwidth]{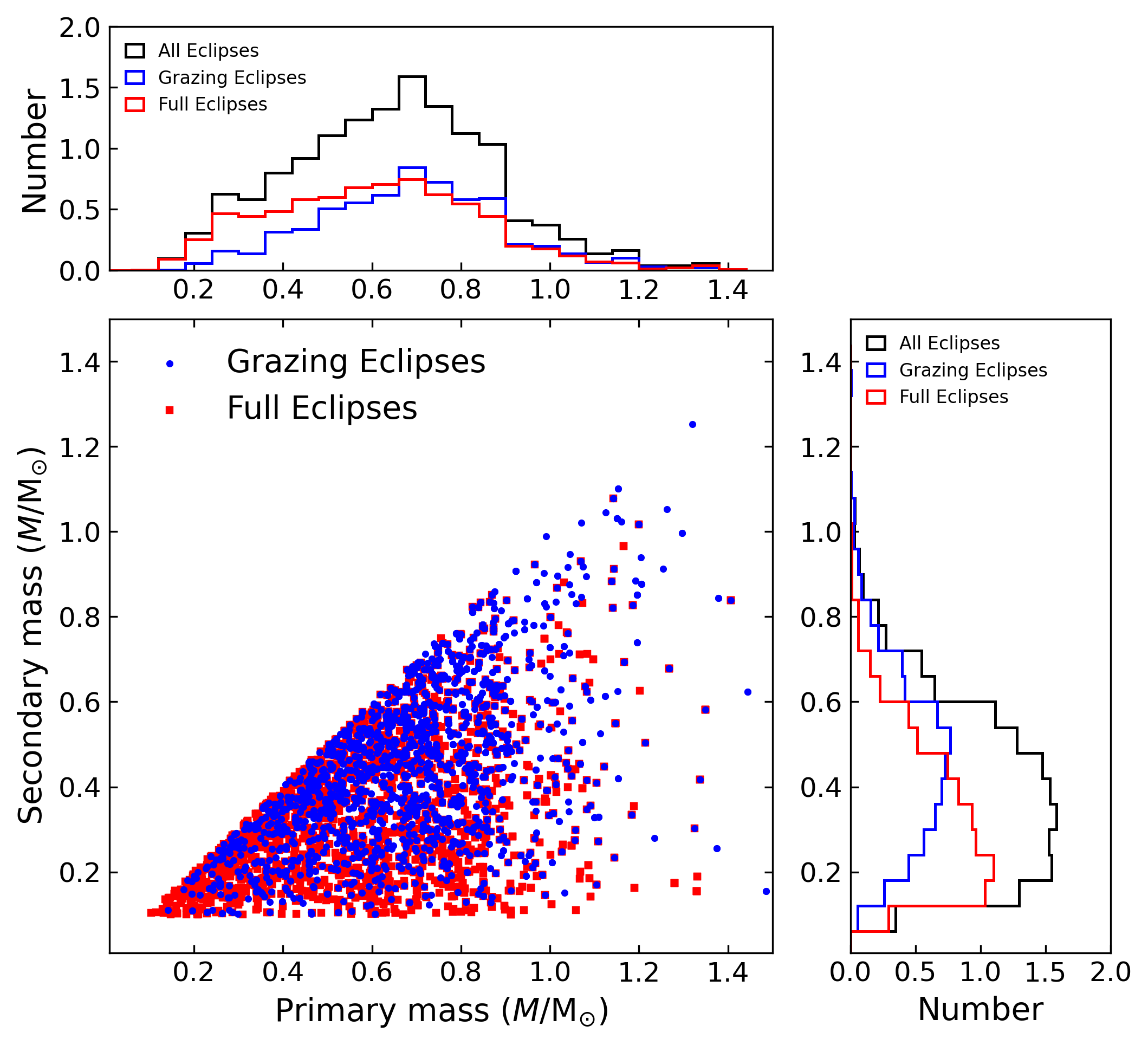}
      \caption{Primary star mass versus secondary star mass by \textbf{eclipse type} for BEBs identified as planet mimics in SYNTH0. Fully eclipsing binaries are shown by red lines in the histogram plots and by red boxes in the scatter plots. Partially eclipsing binaries are shown by blue lines in the histogram plots and blue circles in the scatter plot.}
         \label{fig:m1}
\end{figure}

\begin{figure}
\centering
   \includegraphics[width=0.7\textwidth, height=0.55\textwidth]{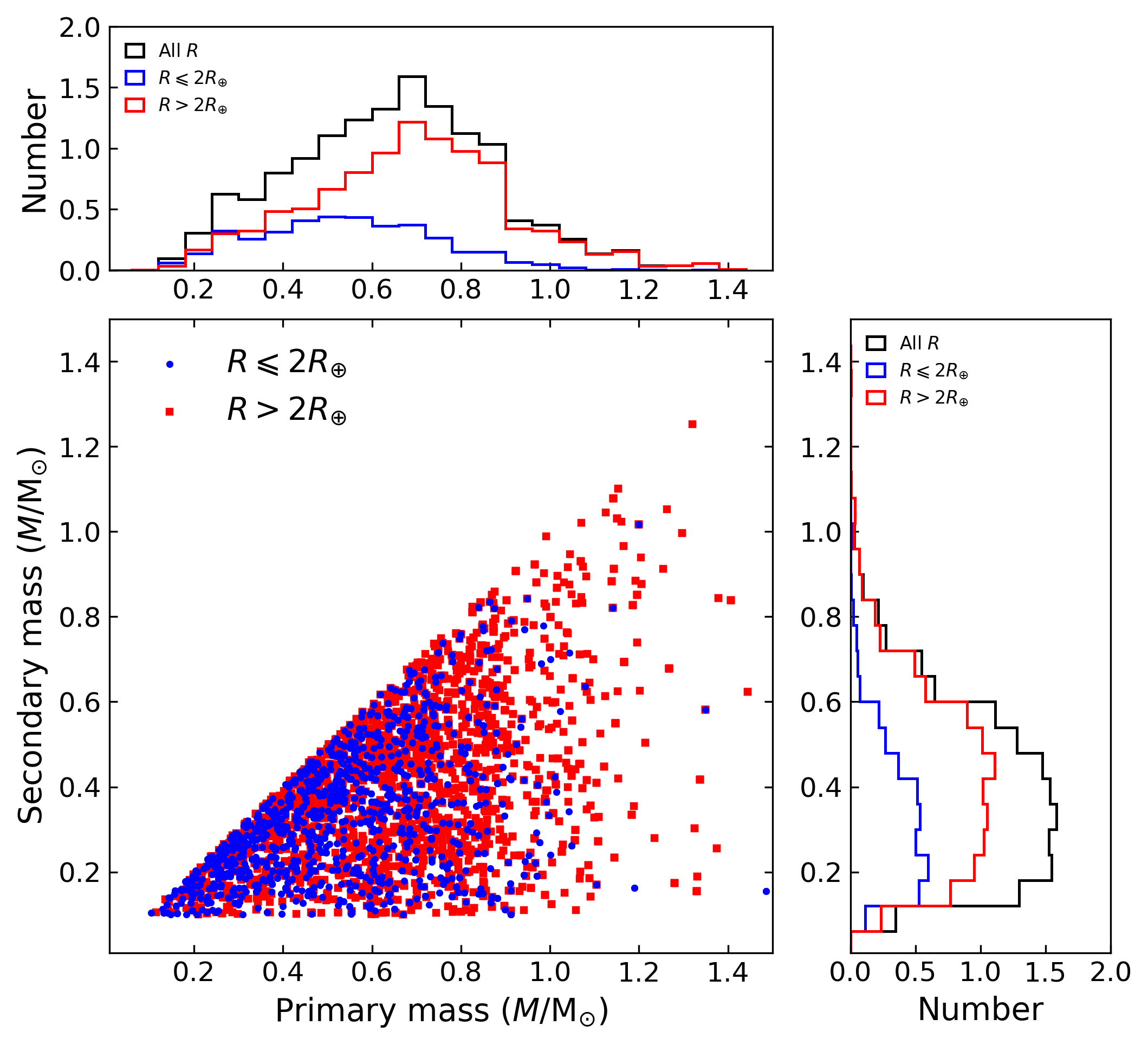}
      \caption{Primary star mass versus secondary star mass by \textbf{apparent planet radius} for BEBs identified as planet mimics in SYNTH0. Mimicked planets with radii log$(R$/R$_{\oplus})>2$ are shown by red lines in the histogram plots and red boxes in the scatter plot. Mimicked planets with radii log$(R$/R$_{\oplus})\leqslant2$ are shown by blue lines in the histogram plots and blue circles in the scatter plot.}
         \label{fig:m2}
\end{figure}

Figure \ref{fig:m1} shows the relationships between the primary star mass ($M_1$) and the secondary mass ($M_2$) by \textbf{eclipse type} while Figure \ref{fig:m2} shows the relationships between the primary star mass ($M_1$) and the secondary mass ($M_2$) by \textbf{planet radius} mimicked. 

In Figure \ref{fig:m1} we see that the distribution of primary masses required for both partially and fully eclipsing binaries to mimic an Earth-like period planet increases continuously with $M_1$ to a maximum at $0.7$M$_{\odot}$ and then drops off gradually until $0.9$M$_{\odot}$ where a sharp drop due to the step function in galactic metallicity occurs. The distribution of secondary masses for partially eclipsing systems shows a broader maximum near $M_2 \approx 0.5$M$_{\odot}$, with a gentle slope to smaller masses and a slightly steeper decrease to larger masses. 
For fully eclipsing binaries the $M_2$ mass distribution shows a broad peak from $0.2$M$_{\odot}$ to $0.5$M$_{\odot}$ with a gentle reduction in slope to larger masses but with a more pronounced drop to smaller masses.

In Figure \ref{fig:m2} we see a similar trend in $M_1$ mass for large radius planet mimics ($R/$R$_{\oplus}>2$), with a gradual increase in $M_1$ up to $\approx 0.7$~M$_{\odot}$ then a decrease until $0.9$M$_{\odot}$ where again the step function in galactic metallicity results in a sharp drop.


For Earth-like period planets with radii $R/$R$_{\oplus} \leqslant2$ we see a broad peak for secondary masses of $M_2 \approx 0.2$M$_{\odot}$ to $0.4$M$_{\odot}$ before a gradual reduction to a maximum secondary binary star mass of $M_2 \approx 0.8$M$_{\odot}$ For large radius planet mimics ($R/$R$_{\oplus}>2$) we see a similar broad maximum at $M_2 \approx 0.2$M$_{\odot}$ to $0.6$M$_{\odot}$ before a gradual reduction to a maximum secondary binary star mass this time extending to a much higher secondary mass of $M_2\approx 1.1$M$_{\odot}$.


While there is a significant reduction in BEBs for $M_1$ masses above $\sim0.9$M$_{\odot}$ which can be seen in the scatter plot of Figure \ref{fig:m1}, both fully and partially eclipsing binaries are observed more or less equally in all areas of the plot. This is in stark contrast to Figure \ref{fig:m2}, where the majority of BEBs with $R/$R$_{\oplus} \leqslant2$ are observed in the region where $M_1 \lessapprox 0.9 $ M$_{\odot}$ and $M_2 \lessapprox 0.8$M$_{\odot}$.

This suggests that primary or secondary stars larger than $M/$M$_{\odot}=1$ are unlikely to produce BEBs for planets in the range $-0.4 \leqslant \log(R/$R$_{\oplus}) \leqslant 1.2$ with Earth-like periods.

\begin{figure}
\centering
   \includegraphics[width=0.7\textwidth, height=0.56\textwidth]{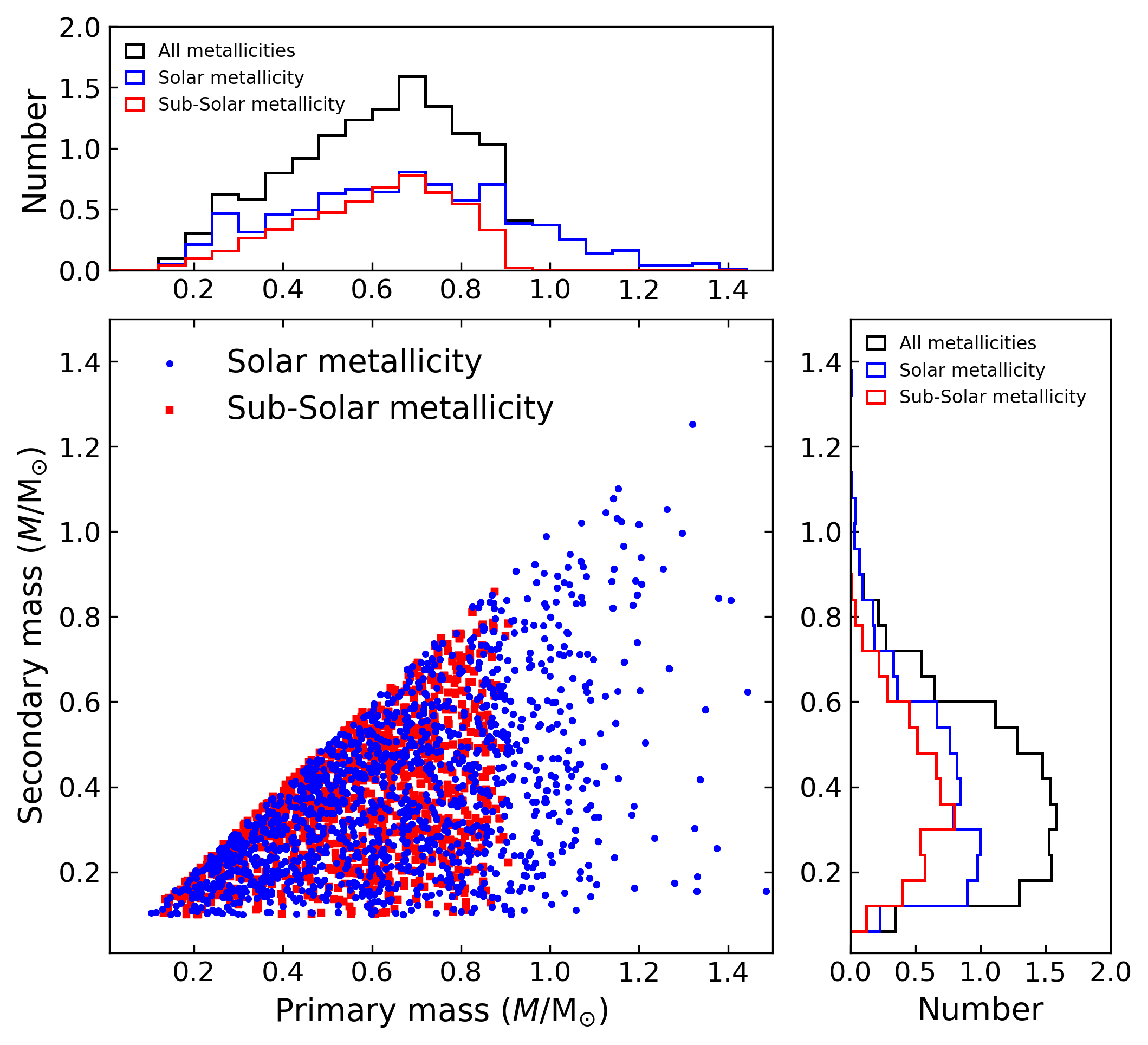}
      \caption{Primary star mass versus secondary star mass by \textbf{metallicity} for BEBs identified as planet mimics in SYNTH0. Mimicked planets with metallicity of $Z=0.0033$ (sub-solar), are shown by red lines in the histogram plots and red boxes in the scatter plot. Mimicked planets with radii metallicity of $Z=0.02$ (solar) are shown by blue lines in the histogram plots and blue circles in the scatter plot.}
        \label{fig:m3}
\end{figure}

In Figure \ref{fig:m3} we see clearly the effect of the metallicity step function used in our galactic model. There are no sub-solar BEBs with primary masses above $M_1 \approx 0.9$M$_{\odot}$ as these stars have all evolved to white dwarfs.

\onecolumn 


\begin{figure}
\centering
    \includegraphics[width=0.65\textwidth, height=0.55\textwidth]
    {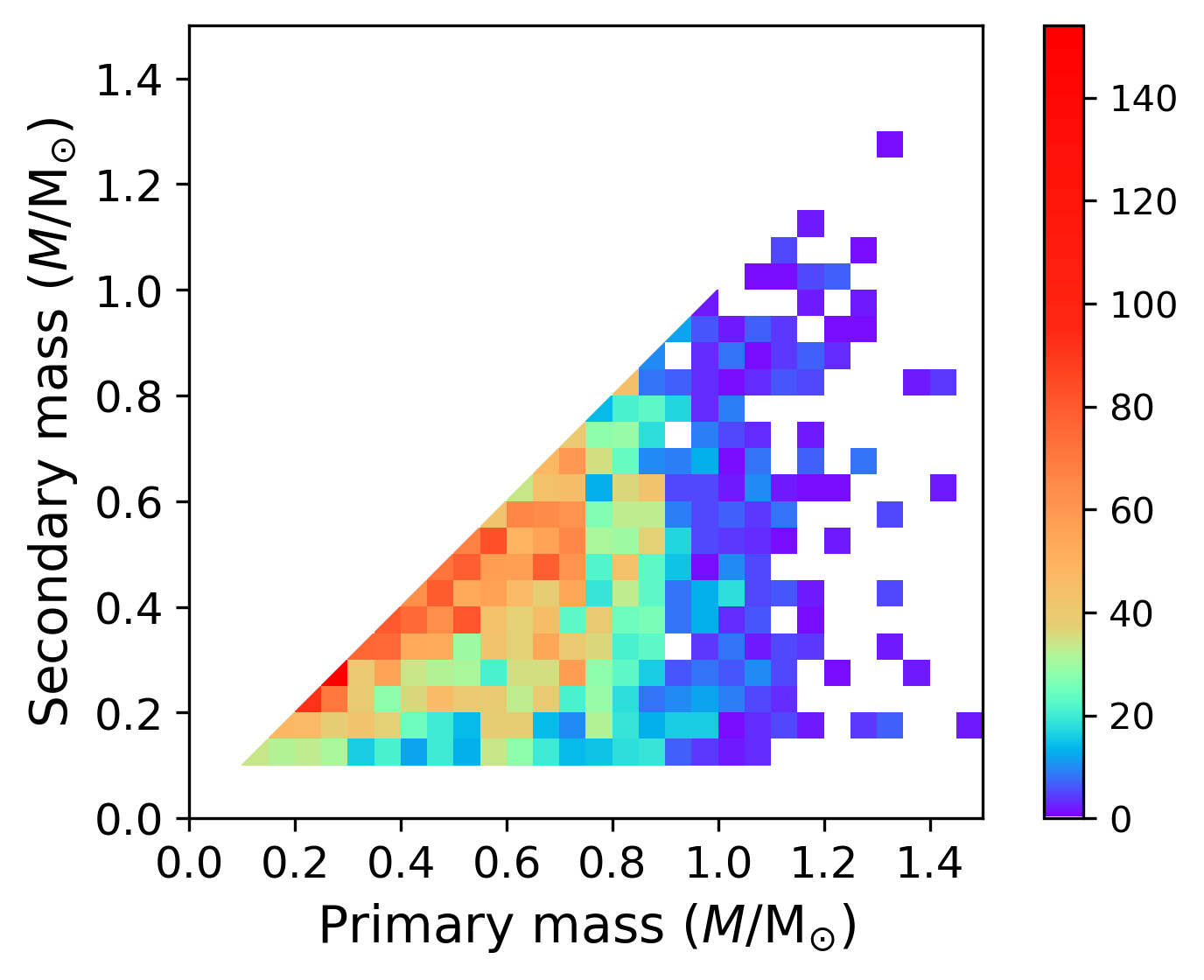}

\caption{Relative frequency of primary and secondary mass combinations resulting in planet mimics in SYNTH0, with radii $-0.4 \leqslant \log(R/$R$_{\oplus}) \leqslant 1.2$ and periods between 180 and 1,000 days. The plot includes all metallicities, eclipse types and apparent planet radii mimicked.}
\label{fig:MassDistrib}
\end{figure}
\twocolumn

\subsection{Relative frequency of primary mass/secondary mass combinations in BEBs}
In Figure \ref{fig:MassDistrib} we show the relative frequency of primary and secondary mass combinations contributing to the BEBs in SYNTH0. Metallicities, eclipse types and apparent planet radii have been combined. There is a clear peak in the primary mass distribution around $0.2<M_1/$M$_{\odot}<0.6$ with the most systems showing nearly equal secondary masses. Notably, very few BEBs come from stars with masses greater than $M_1\approx1.0$M$_{\odot}$ although we once again caution that while we believe this feature is clearly evident, it is likely amplified by the metallicity step function we have used in modeling the Galaxy.

\subsection{Planet sizes mimicked by BEBs}
\begin{figure}
\centering
   \includegraphics[width=\hsize]{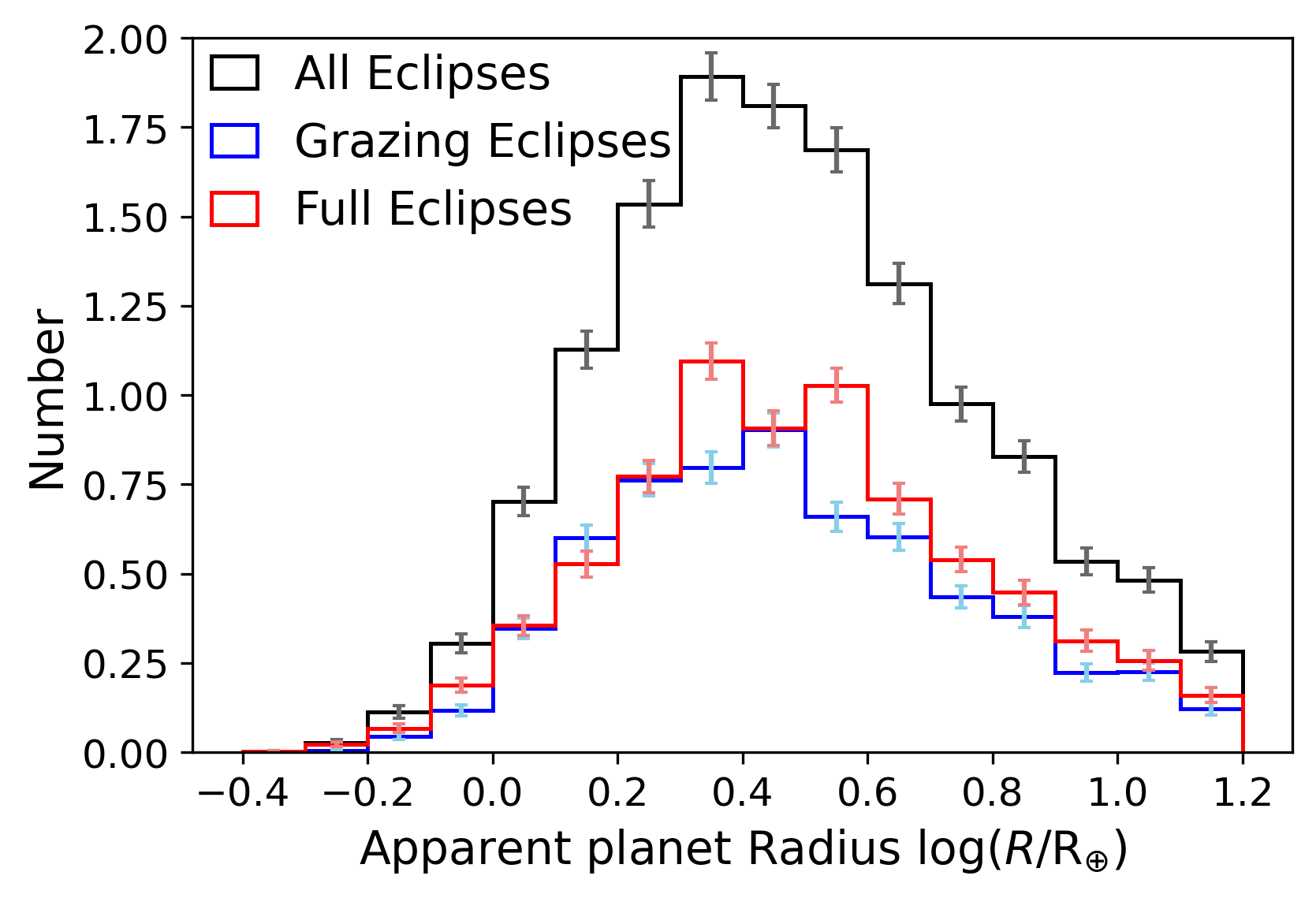}
      \caption{Number of binary eclipses in SYNTH0 by eclipse type, mimicking the transits of planets with apparent radii $-0.4 \leqslant \log(R/$R$_{\oplus}) \leqslant 1.2$ and periods between 180 and 1,000 days. Numbers are averaged over 400 simulation runs and include full and grazing eclipses as outlined in Section \ref{seba}.}
         \label{fig:mimicET}
\end{figure}

In Figure \ref{fig:mimicET} we show the number of binary eclipses in SYNTH0 mimicking planets with radii $-0.4 \leqslant \log(R/$R$_{\oplus}) \leqslant 1.2$ and periods between  180 and 1,000 days by eclipse type. There is an apparent maximum in the radius range at $\log(R/$R$_{\oplus}) \approx 0.3-0.6$ with both grazing and full eclipses peaking in this range. This is well above the radius range of interest to the PLATO mission. The very low number of BEBs in the range $\log(R/$R$_{\oplus}) \leqslant 0$ must be interpreted with caution. Our analysis in Paper 1 \cite[refer to Figures 21 and 22 in][]{Bray_2023}, shows that very few BEBs (or PTs), are detected in the range $\log(R/$R$_{\oplus}) \leqslant 0$. Therefore, the reduction of BEBs with apparent radii $\log(R/$R$_{\oplus}) \leqslant 0$ is most likely a result of PLATO's detector characteristics rather than some property of the surrounding stellar population.

\subsection{Location and distances to BEBs}
We analysed the BEB locations by eclipse type and planet radius bin for all 400 simulation runs. We find a higher number of BEBs at latitudes closer to the Galactic plane and a slightly increased number at increased longitudes. As expected, the number of BEBs drops significantly as we move away from the Galactic plane but otherwise both distributions are unremarkable with no significant features other than those imprinted by the extinction map. 

\begin{figure}
\centering
   \includegraphics[width=\hsize]{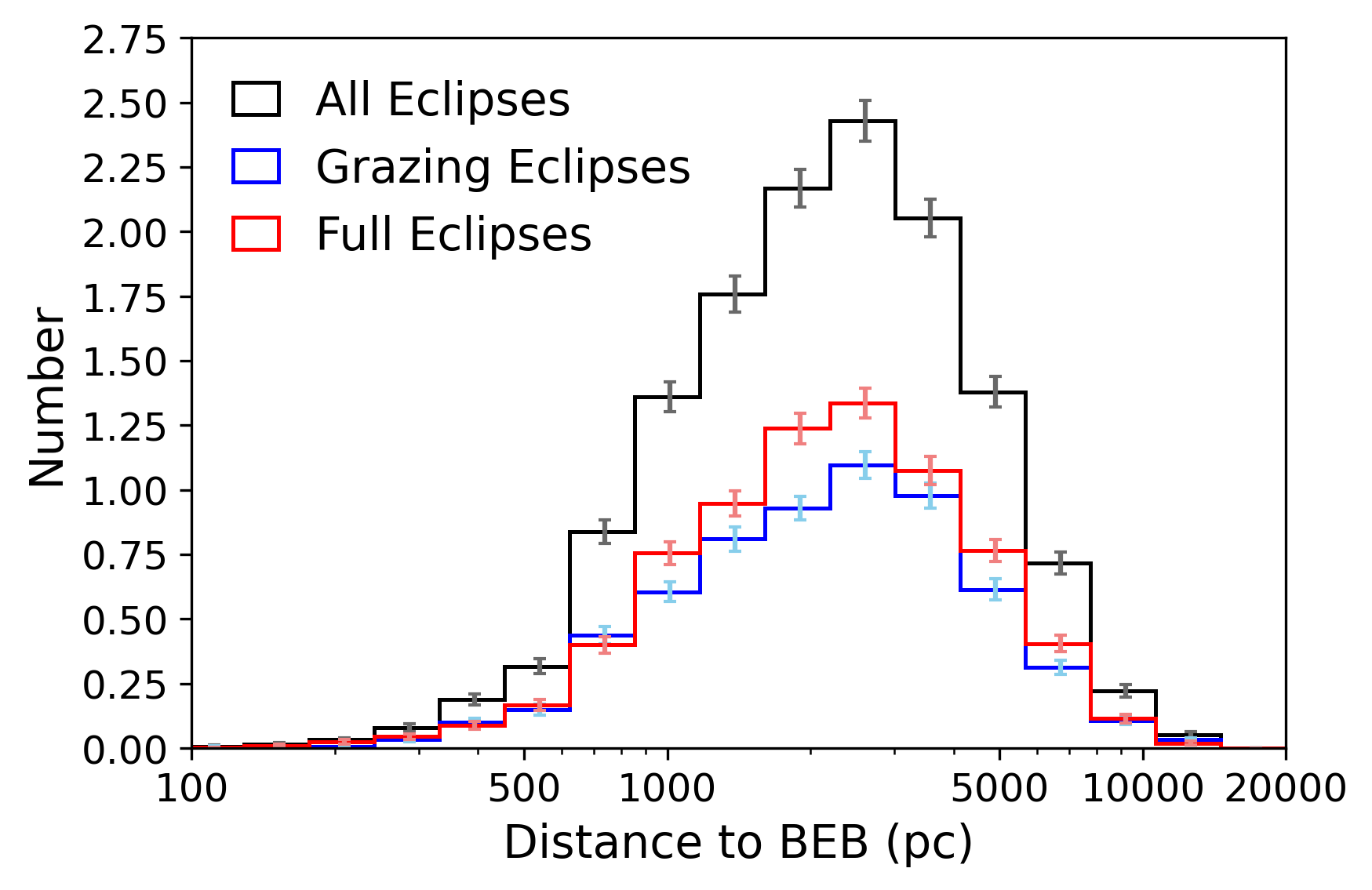}
      \caption{Distance to binary eclipses in SYNTH0 by eclipse type, mimicking the transits of planets with apparent radii $-0.4 \leqslant \log(R/$R$_{\oplus}) \leqslant 1.2$ and periods between 180 and 1,000 days. Numbers are averaged over 400 simulation runs and include full and grazing eclipses as outlined in Section \ref{seba}.}
         \label{fig:distET}
\end{figure}

\begin{figure}
\centering
   \includegraphics[width=\hsize]{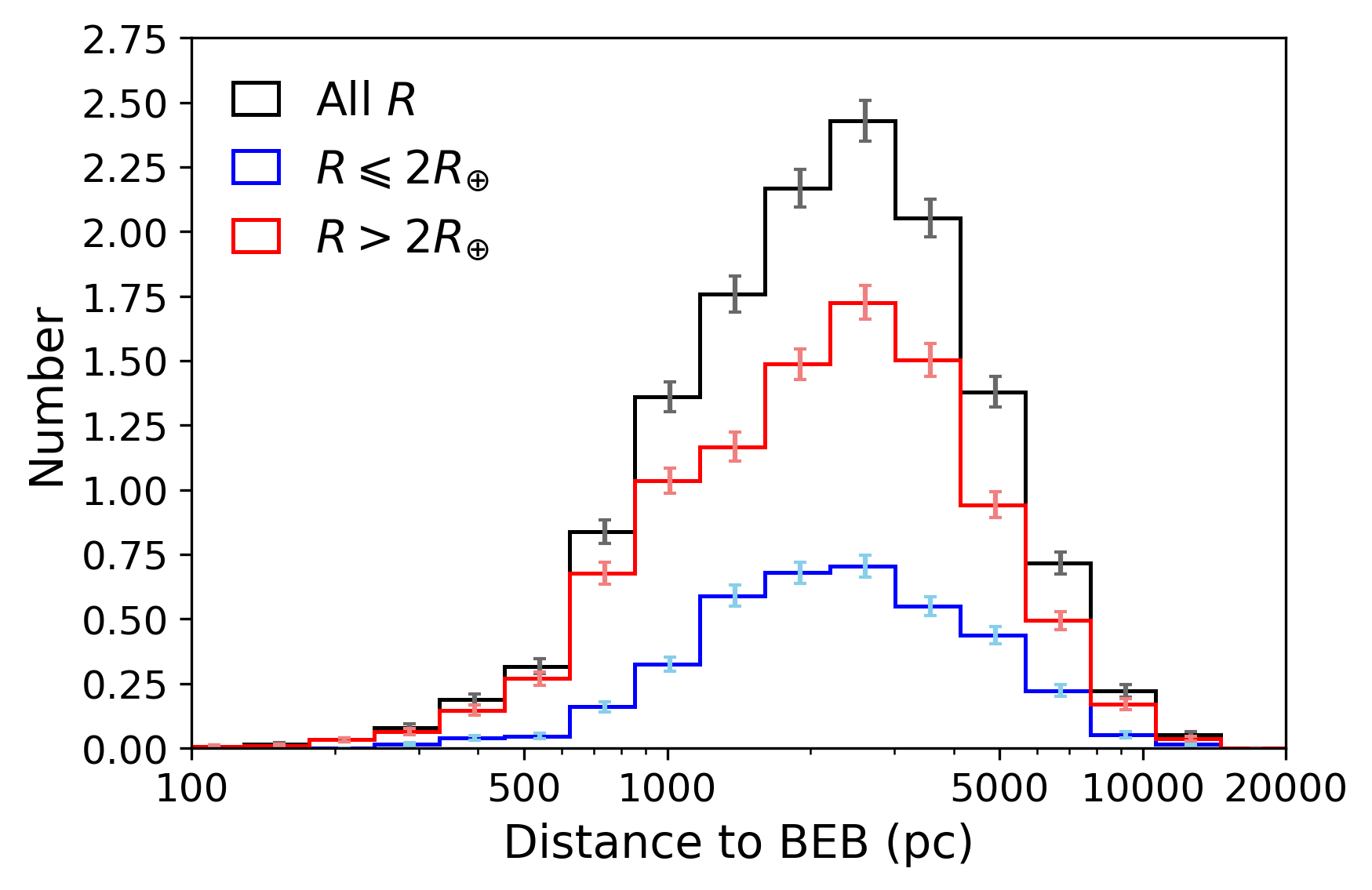}
      \caption{Distance to binary eclipses in SYNTH0 by apparent radius, mimicking the transits of planets with apparent radii $-0.4 \leqslant \log(R/$R$_{\oplus}) \leqslant 1.2$ and periods between 180 and 1,000 days. Numbers are averaged over 400 simulation runs and include full and grazing eclipses as outlined in Section \ref{seba}.}
         \label{fig:distR}
\end{figure}

Distances to the binary contaminants by eclipse type are shown in Figure \ref{fig:distET} and by planet radius group in Figure \ref{fig:distR}. While it is difficult to identify an obvious trend, from these figures it would appear that there is no overall difference in the distances to the BEBs for the planet radius groups or the eclipse type, with both appearing to peak at the same distance of $\sim2,000 - 3,000$pc. Smaller planets seem to display a wider peak than larger planets around this distance but otherwise all of the distributions are very similar. As expected, the number of BEBs reduces rapidly as the distance increases and the background binaries become too faint to mimic any planet transits.

\subsection{Radius ratio of BEBs}\label{rad}

\begin{figure}
\centering
   \includegraphics[width=\hsize]{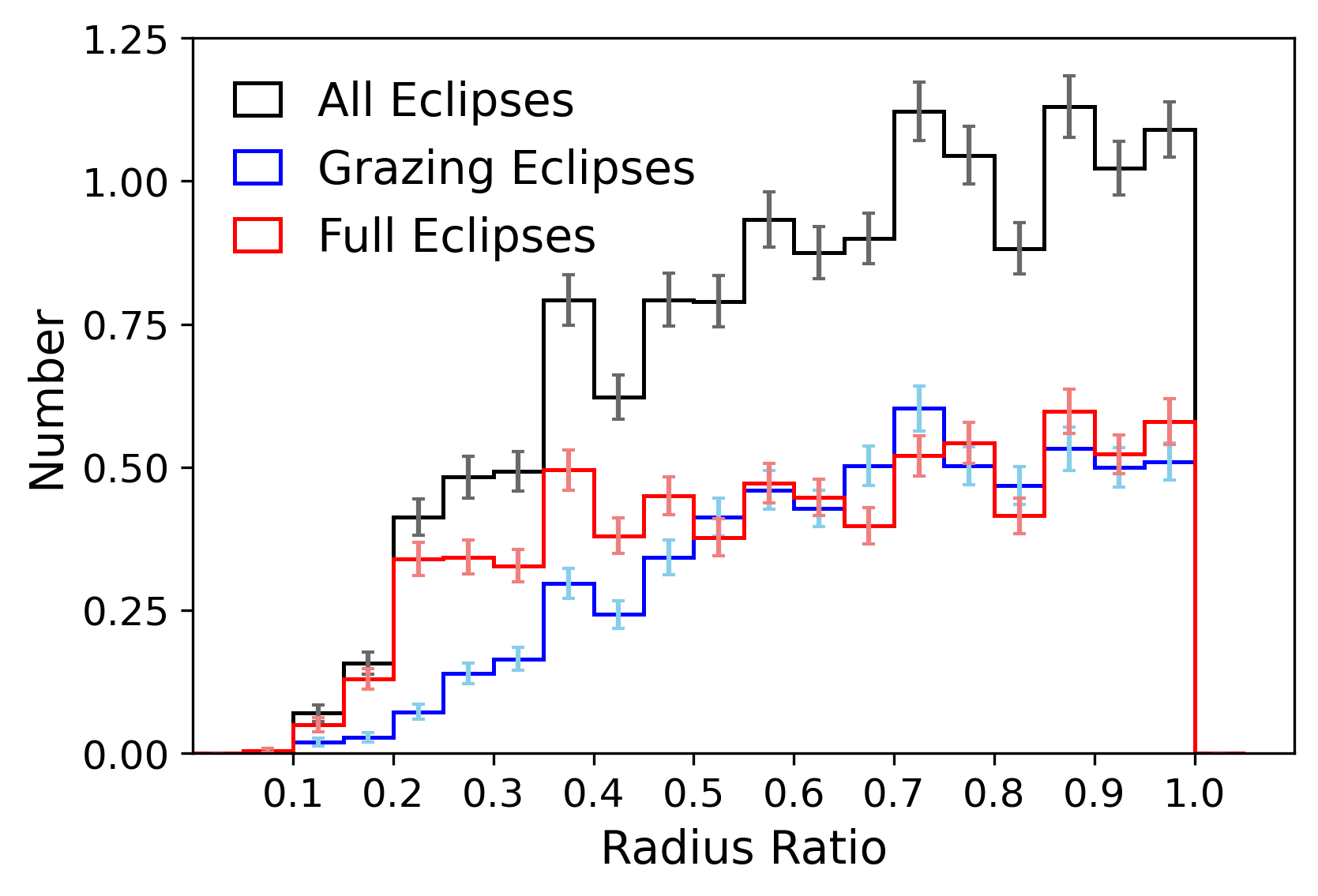}
      \caption{Radius ratio of binary eclipses in SYNTH0 shown by eclipse type, mimicking the transits of planets with apparent radii $-0.4 \leqslant \log(R/$R$_{\oplus}) \leqslant 1.2$ and periods between 180 and 1,000 days. Numbers are averaged over 400 simulation runs and include full and grazing eclipses as outlined in Section \ref{seba}.}
         \label{fig:radratET}
\end{figure}

\begin{figure}
\centering
   \includegraphics[width=\hsize]{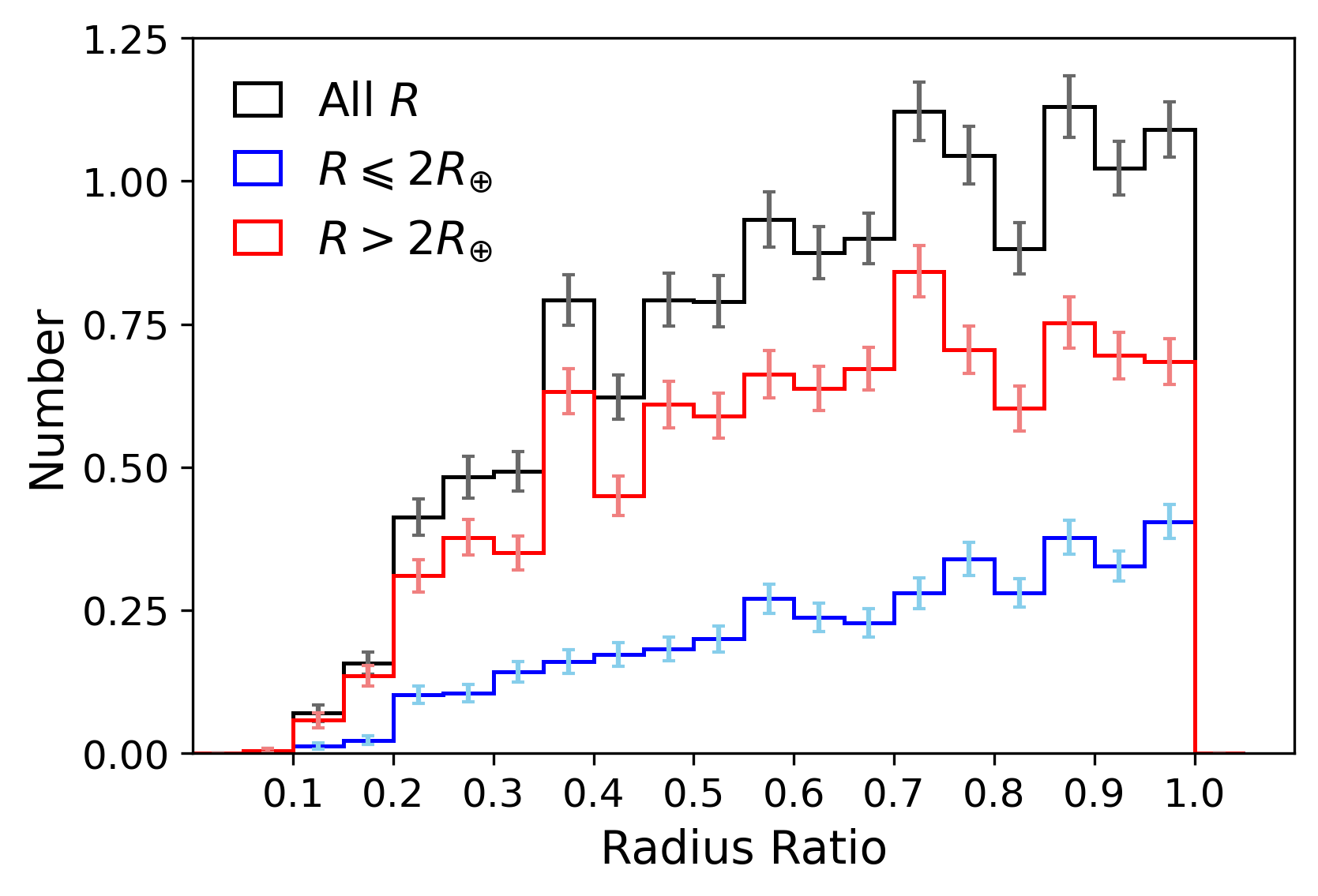}
      \caption{Radius ratio of binary eclipses in SYNTH0 shown by apparent radius, mimicking the transits of planets with apparent radii $-0.4 \leqslant \log(R/$R$_{\oplus}) \leqslant 1.2$ and periods between 180 and 1,000 days. Numbers are averaged over 400 simulation runs and include full and grazing eclipses as outlined in Section \ref{seba}.}
         \label{fig:radratR}
\end{figure}

Figures \ref{fig:radratET} and \ref{fig:radratR} show the distribution of radius ratio (secondary star radius divided by primary star radius) of our contaminating binaries. The radius ratio largely reflects the mass ratio of the binary. 

Figure \ref{fig:radratET} while somewhat noisey, suggests that that for grazing eclipses we see more BEBs as the radius ratio approaches 1. The closer the radius ratio is to 1, the bigger the secondary is relative to the primary, and the lower the angle required for a grazing eclipse is. Hence, the grazing eclipse probability increases with increasing radius ratio.

For fully eclipsing binaries the trend is more complex. Increasing the secondary radius relative to the primary radius will  increase the critical inclination angle for a full eclipse, making it less likely, however countering this effect is the eclipse depth. With an increasing radius ratio the full eclipse becomes deeper, and so is still detectable even when the magnitude difference between target star and background blend is larger. As a result of these competing effects full eclipses follow the same trend as grazing eclipses albeit for different reasons.

Considering the radius ratio for eclipses by planet mimic radii shown in Figure \ref{fig:radratR}, we find that for large planet radii there is a gradual increase in the number of BEBs mimicking larger planets as the radius of the secondary star increases and approaches the radius of the primary star. This is again expected as the increased area eclipsed would mimic a larger planet. 



For smaller planet radii we observe a similar trend as the radius of the secondary star increases and approaches the radius of the primary star. This is again as expected, because as the radius of the secondary star increases, the critical angle for a grazing eclipse reduces, resulting in more grazing eclipses.



\subsection{Period distribution of BEBs}
\begin{figure}
\centering
   \includegraphics[width=\hsize]{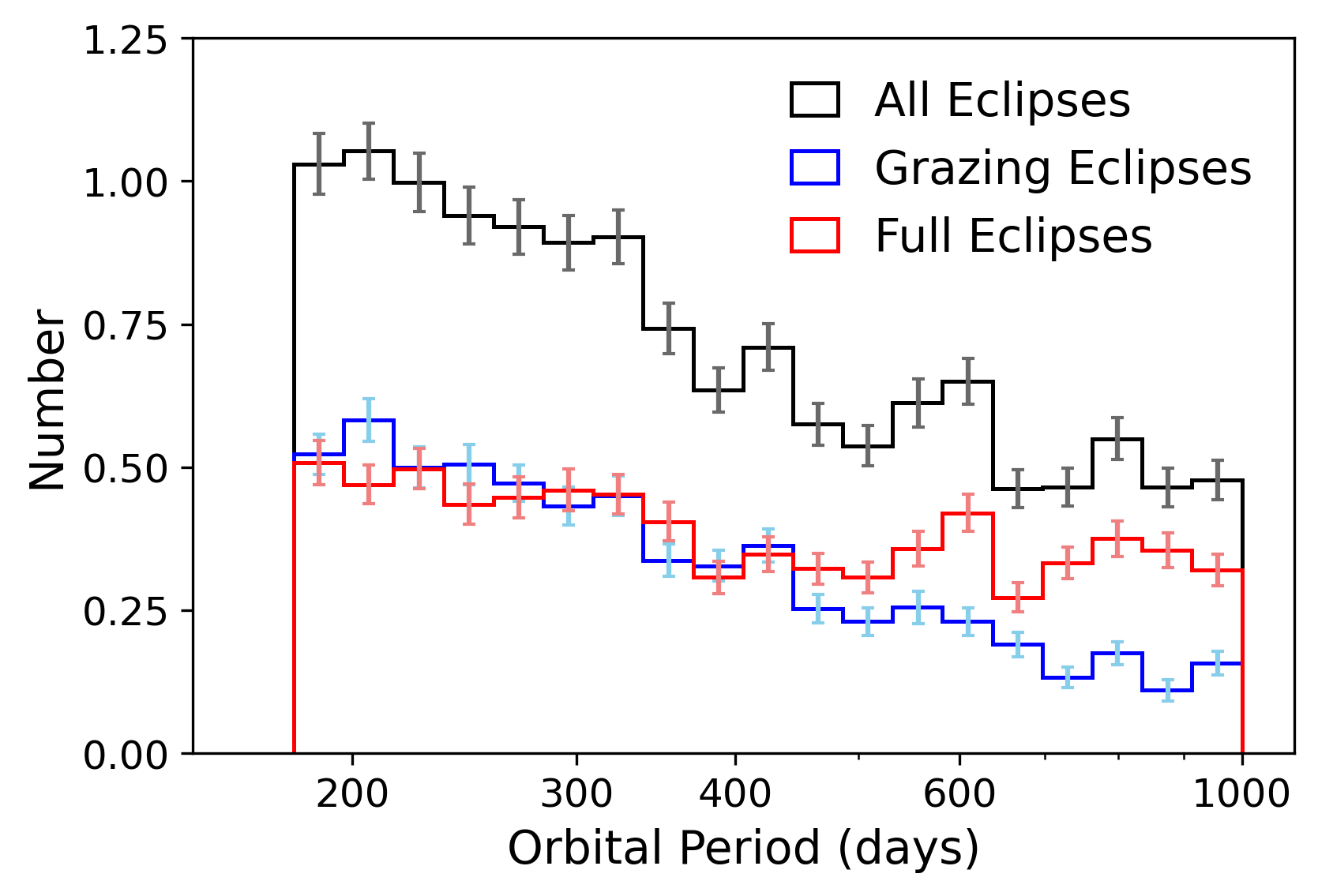}
      \caption{Orbital periods of binary eclipses in SYNTH0 shown by eclipse type, mimicking the transits of planets with apparent radii $-0.4 \leqslant \log(R/$R$_{\oplus}) \leqslant 1.2$ and periods between 180 and 1,000 days. Numbers are averaged over 400 simulation runs and include full and grazing eclipses as outlined in Section \ref{seba}. Note the orbital period scale is logarithmic.}
         \label{fig:porbET}
\end{figure}

Figure \ref{fig:porbET} shows the period distribution by eclipse type, for periods between 180 to 1,000 days. The incidence of eclipses reduces slightly as the period increases since the critical inclination angle moves closer to $90^{\circ}$. 

\begin{figure}
\centering
   \includegraphics[width=\hsize]{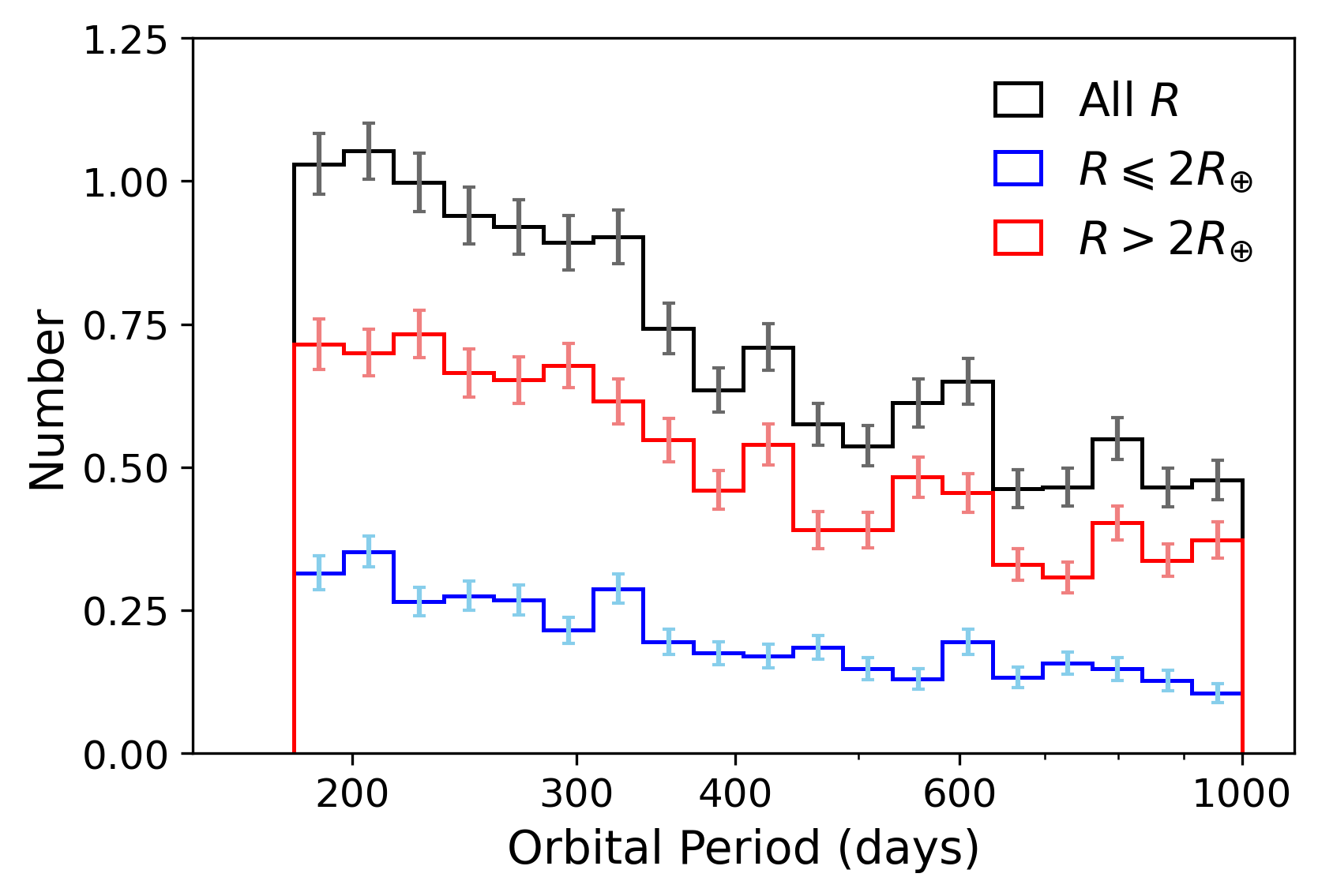}
      \caption{Orbital periods of binary eclipses in SYNTH0 shown by apparent radius, mimicking the transits of planets with apparent radii $-0.4 \leqslant \log(R/$R$_{\oplus}) \leqslant 1.2$ and periods between 180 and 1,000 days. Numbers are averaged over 400 simulation runs and include full and grazing eclipses as outlined in Section \ref{seba}. Note the orbital period scale is logarithmic.}
         \label{fig:porbR}
\end{figure}

The orbital period distribution separated by planet mimic radius is displayed in Figure \ref{fig:porbR}. The distribution of large radius mimics shows a clear increase towards shorter periods flattening off slightly below 300 days, a trend that is much less noticeable for small radius mimic planets. As orbital periods decrease, deeper eclipses, mimicking larger radii planets, are more likely.

\subsection{Orbital inclination angle distribution of BEBs}

\begin{figure}
\centering
   \includegraphics[width=\hsize]{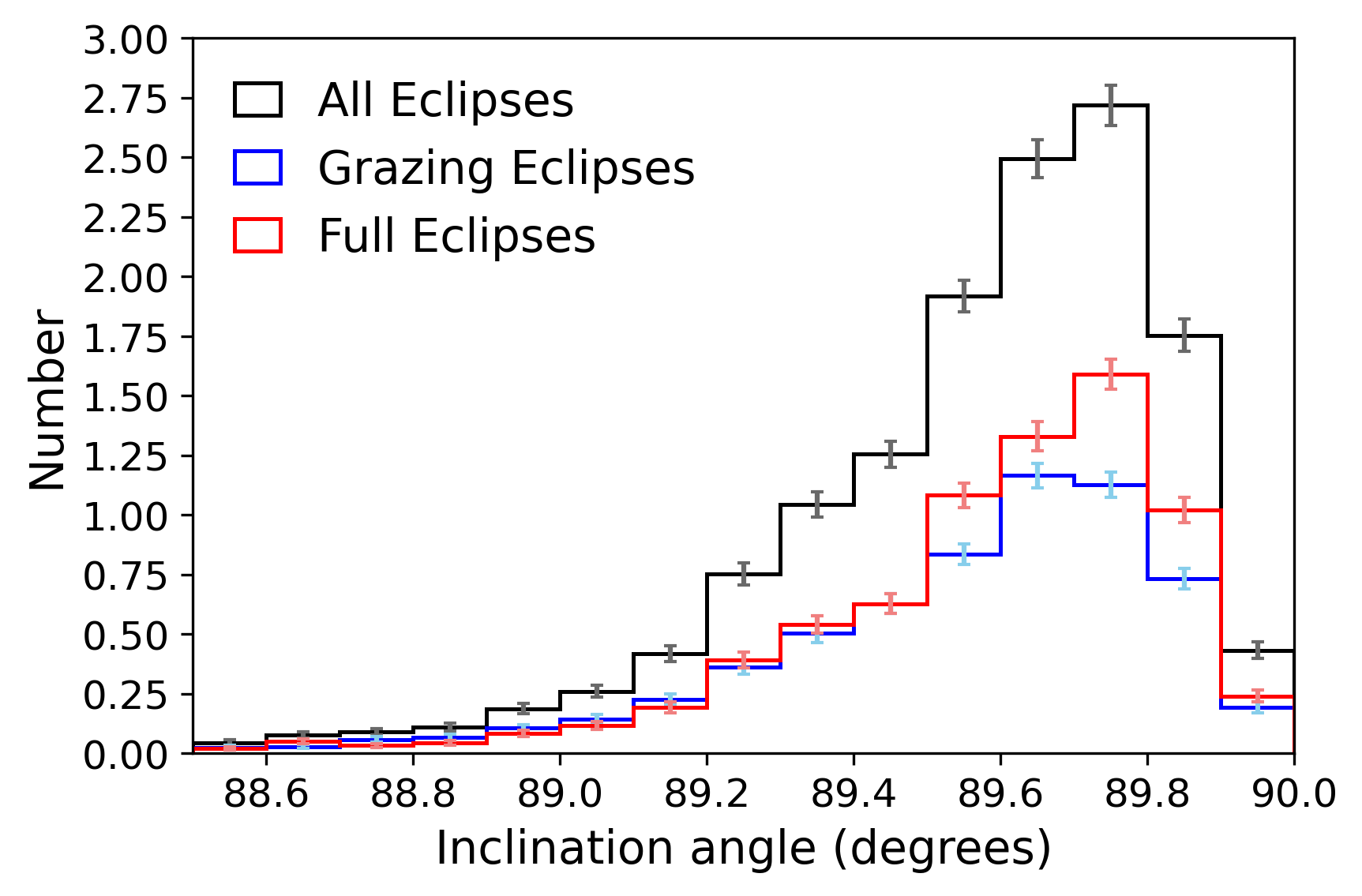}
      \caption{The implied orbital inclinations of the planets mimicked by binary eclipses in SYNTH0 binned by eclipse type, mimicking the transits of planets with apparent radii $-0.4 \leqslant \log(R/$R$_{\oplus}) \leqslant 1.2$ and periods between 180 and 1,000 days. Numbers are averaged over 400 simulation runs and include full and grazing eclipses as outlined in Section \ref{seba}.}
         \label{fig:incPLET}
\end{figure}

\begin{figure}
\centering
   \includegraphics[width=\hsize]{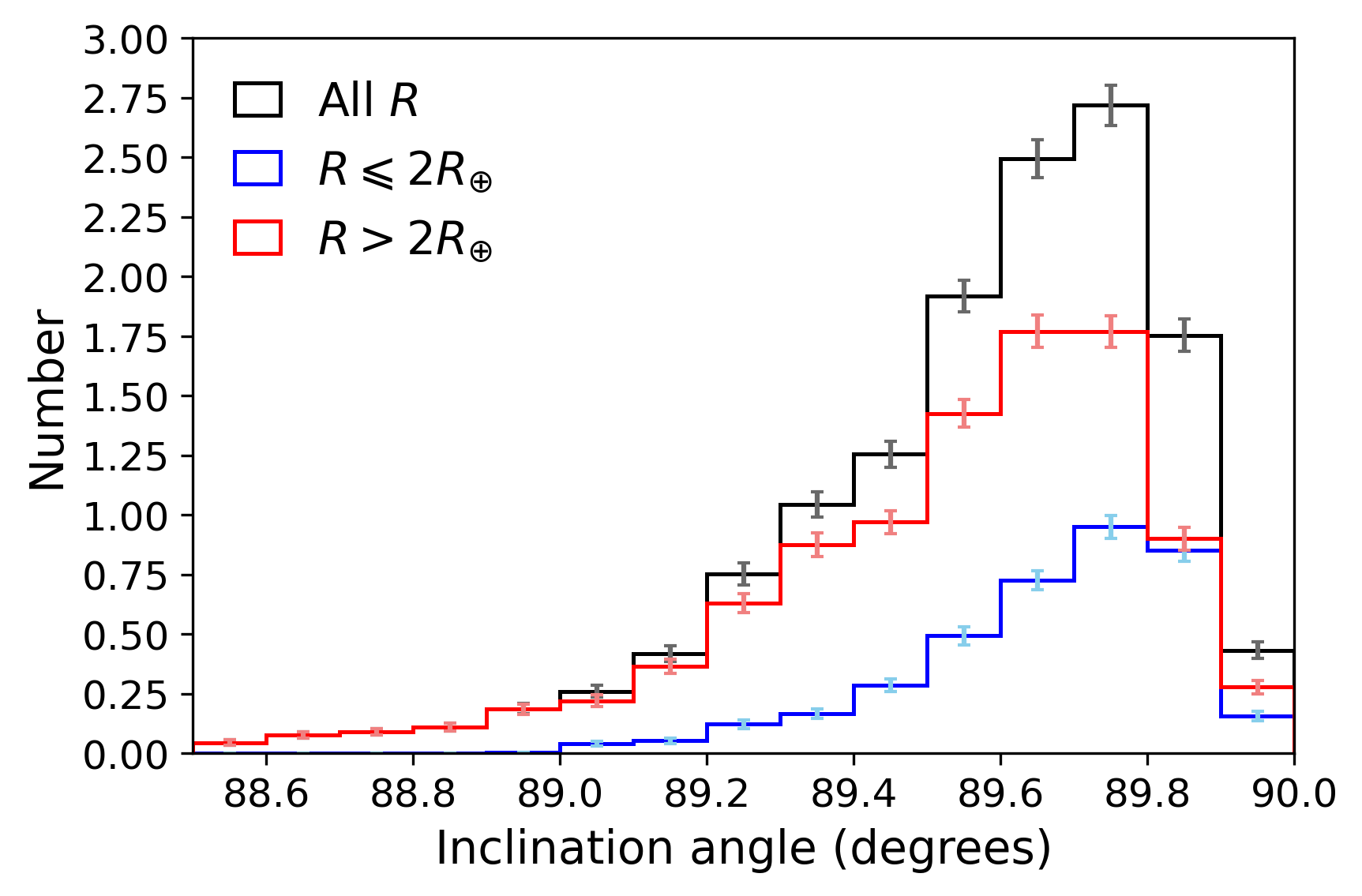}
      \caption{The implied orbital inclinations of the planets mimicked by binary eclipses in SYNTH0 binned by planet radius, mimicking the transits of planets with apparent radii $-0.4 \leqslant \log(R/$R$_{\oplus}) \leqslant 1.2$ and periods between 180 and 1,000 days. Numbers are averaged over 400 simulation runs and include full and grazing eclipses as outlined in Section \ref{seba}.}
         \label{fig:incPLR}
\end{figure}

In Figures \ref{fig:incPLET} and \ref{fig:incPLR} we show the implied orbital inclinations of the planets mimicked by the BEBs which are consistent with the target star and apparent planet radius pairings. 
To ensure consistency we use Equation \ref{Eqn22} to calculate the mimicked inclination angle using the BEB eclipse duration, the target star radius and mimicked planet radius. We then calculate the critical inclination angle for the mimicked planet radius and target radius pairings. Only inclination angles calculated in Equation \ref{Eqn22} that are greater than the corresponding critical inclination angles are included.

The BEB inclination angle distributions are largely as expected, with full eclipses and larger radii planets requiring inclination angles closer to $90^{\circ}$ than grazing eclipses and planets mimics of smaller radii. 
However, we also find that the inclination angle distribution peaks at approximately $89.7^{\circ}$ not the expected $90^{\circ}$, for both planet radii and eclipse type inclination angle distributions. For eclipsing binary systems with inclinations very close to $90^{\circ}$, the eclipse duration is too long for the corresponding planet radii calculated from the flux drop and these BEBs are removed by the transit duration check. 

We reiterate that the planet inclination angle distributions shown in Figures \ref{fig:incPLET} and \ref{fig:incPLR} are `fake' inclinations calculated from the observed eclipse depths and durations. The test used in Section \ref{gpebc1} calculates what the minimum inclination angle would be based on the `apparent' planet radius and eliminates those BEBs whose calculated inclination angles violate Equation \ref{Eqn22} for the transit duration.

\subsection{Effect of new LOPS field location on our results}
As mentioned in Section \ref{tpm}, the final location for the LOPS field has been confirmed and centred on $l \approx 255.94^{\circ}$ and $b \approx -24.62^{\circ}$. This represents a shift of $\approx 5.6^{\circ}$ in longitude closer to the Galactic centre and a shift of $\approx 5.4^{\circ}$ degrees latitude closer to the Galactic plane. While the effect of the longitude change is expected to be minimal (see Paper 1), the latitude change drops part of the field into the plane itself and into a significantly more dense stellar environment. This is evident from the revised PIC which shows an increase in target star numbers of $\approx16$ percent (140,105 PIC110 to 167,103 PIC200), and we expect a similar magnitude increase in both the PT and BEB numbers. While the number of BEBs and PTs will increase, we expect no significant change in either the PT or BEB distributions. We suggest that a $\approx16$ percent increase on the numbers provided in this analysis will give a reasonable estimate for the numbers expected from the revised LOPS2 location.

\section{Discussion and conclusions}\label{dc}
In this paper we have expanded on the research in \cite{Bray_2023} and attempted to identify the statistical properties of the BEBs expected to be observed by the PLATO mission in the hope that this information will be useful in identifying and eliminating potential stellar groups which will result in BEB planetary signals. 

After verifying that the target star population in SYNTH0 provides a reasonable approximation of PICtarget110 (and PICtarget200), we analysed 400 independent simulation runs of the entire initially proposed  Southern PLATO field (LOPS1),  to identify statistical trends in the resulting BEB population.


We have introduced an additional check to ensure that the BEB duration and flux drop from the BEB binary eclipse simulated could be generated by the target star / planet radius combination inferred. This results in approximately 30 percent of our original BEBs being discounted as no planet orbital inclination angle can produce such events.

We have analysed, in detail, all BEBs for planets of radii $-0.4 \leqslant \log(R/$R$_{\oplus}) \leqslant 1.2$ in Earth-like periods (which we define as between 180 and 1,000 days).

We note that the step function used in our metallicity mix (either $Z=0.02$ or $Z=0.0033$), has resulted in some artificial discontinuities, most notably in the absolute exclusion of sub-solar metallicity binaries with a primary mass greater than $\approx0.9$M$_{\odot}$ in any BEBs.

Our main findings are as follows;
\begin{itemize}
    \item [$\bullet$] BEBs in the radius range analysed are not likely to be generated by binaries where any stellar component is greater than $1$M$_{\odot}$
    \item [$\bullet$] BEBs for planets with $R/$R$_{\oplus} \leqslant2$ will most likely be generated by low mass binaries where the mass of each binary star component is $\lessapprox0.5$M$_{\odot}$  
    \item [$\bullet$] BEBs for planets with $R/$R$_{\oplus} >2$ are most likely to be generated from binaries with primary masses around $0.7$M$_{\odot}$ and smaller mass secondaries in the range $0.2 \lessapprox M/$M$_{\odot}\lessapprox0.6$.
    \item [$\bullet$] Solar metallicity binaries contribute to BEBs almost equally throughout the 0.2 to $1.2$M$_{\odot}$ primary stellar mass range. But since there are virtually no sub-solar metallicity primary stars greater than $\approx0.9$M$_{\odot}$ only sub-solar metallicity primary stars less than $\approx0.9$M$_{\odot}$ contribute to BEBs.
    \item [$\bullet$] There appears to be no identifiable trend by planet radius or eclipse types with respect to distance to the BEBs or radii of the mimicked planet.
    \item [$\bullet$] We find, as expected, that the peak for BEBs by radius ratio occurs close to one for both full and grazing eclipses as a relatively larger secondary will eclipse its primary star already for a smaller inclination angle. However, the distribution flattens as the radius ratio limit of one is approached.
    \item [$\bullet$] 
    The period distribution for planet mimics in the 180 - 1,000 day period range generally increases towards shorter periods, but this trend is much more pronounced for planets with planet mimic radii larger than $R/$R$_{\oplus}\approx 2$. 
    \item [$\bullet$] As expected, the number of both grazing and fully eclipsing binaries resulting in BEBs increases as the implied orbital inclinations of the planets mimicked
    approaches $90^{\circ}$. However, 
    the peak in this inclination distribution occurs around $89.7^{\circ}$ rather than at $90^{\circ}$. This is because as the inclination angle increases so too does the transit time, and close to $90^{\circ}$ a point is reached when the transit time is too long for the inferred planet/target star combination.
    
\end{itemize}

Future follow-up on candidates that eventually turn out to be BEBs will be extremely useful in verifying the validity of the models used. More generally, comparing model predictions such as the ones presented here with the collective properties of verified BEBs that will emerge during the operational phase of PLATO will test the assumptions made for the physics of stellar and binary formation and evolution.




\section*{Acknowledgements}

We would like to thank the anonymous referee for their thoughtful analysis and insightful suggestions to improve the paper.

This work presents results from the European Space Agency (ESA) space mission PLATO. The PLATO payload, the PLATO Ground Segment and PLATO data processing are joint developments of ESA and the PLATO Mission Consortium (PMC). Funding for the PMC is provided at national levels, in particular by countries participating in the PLATO Multilateral Agreement (Austria, Belgium, Czech Republic, Denmark, France, Germany, Italy, Netherlands, Portugal, Spain, Sweden, Switzerland, Norway, and United Kingdom) and institutions from Brazil. Members of the PLATO Consortium can be found at https://platomission.com/. The ESA PLATO mission website is https://www.cosmos.esa.int/plato. We thank the teams working for PLATO for all their work.\\
\\
JCB acknowledges the support provided by the University of Auckland and funding from the Royal Society Te Ap$\bar{\textrm{a}}$rangi of New Zealand Marsden Grant Scheme.\\
\\
UK acknowledges support by STFC grant  ST/T000295/1.\\ 
\\
This research was supported by UKSA grant ST/R003211/1 (Open University element of PLATO UK - Support for the Development Phase)\\

\section*{Data Availability}

The data used in this research is available on request from the authors.



\bibliographystyle{mnras}
\bibliography{PLATO.bib}








\bsp	
\label{lastpage}
\end{document}